\documentclass[onecolumn,showpacs,aps,prd]{revtex4}
\usepackage{graphicx,subfig,rotate,color}
\usepackage{amsmath,amssymb,latexsym}
\usepackage{graphicx}
\usepackage{dcolumn}
\usepackage{amsmath}
\usepackage{epsfig}
\RequirePackage{xspace}
\begin{document}
\def\arccosh{{\rm arccosh}}
\def\OL{\overline} 
\def\bp{{b^\prime}}
\def\tp{{t^\prime}}
\def\acpt{\alpha}
\def\op{{\cal O}}
\def\lsim{\mathrel{\lower4pt\hbox{$\sim$}}\hskip-12pt\raise1.6pt\hbox{$<$}\;}
\def\Dd{\psi}
\def\pp{\lambda}
\def\ket{\rangle}
\def\BAR{\bar}
\def\xba{\bar}
\def\fa{{\cal A}}
\def\fm{{\cal M}}
\def\fl{{\cal L}}
\def\ufs{\Upsilon(5S)}
\def\gsim{\mathrel{\lower4pt\hbox{$\sim$}}
\hskip-10pt\raise1.6pt\hbox{$>$}\;}
\def\ufour{\Upsilon(4S)}
\def\xcp{X_{CP}}
\def\ynotcp{Y}
\vspace*{-.5in}
\def\ETAp{\ETA^\prime}
\def\bfb{{\bf B}}
\def\fd{r_D}
\def\fb{r_B}
\def\ed{\ETA_D}
\def\eb{\ETA_B}
\def\hatA{\hat A}
\def\hatfd{{\hat r}_D}
\def\hated{{\hat\ETA}_D}
\def\D{{\bf D}}
\def\pcc{(+ charge conjugate)}

\def\uglu{\hskip 0pt plus 1fil
minus 1fil} \def\uglux{\hskip 0pt plus .75fil minus .75fil}

\def\slashed#1{\setbox200=\hbox{$ #1 $}
    \hbox{\box200 \hskip -\wd200 \hbox to \wd200 {\uglu $/$ \uglux}}}

\def\slpar{\slashed\partial}
\def\sla{\slashed a}
\def\slb{\slashed b}
\def\slc{\slashed c}
\def\sld{\slashed d}
\def\sle{\slashed e}
\def\slf{\slashed f}
\def\slg{\slashed g}
\def\slh{\slashed h}
\def\sli{\slashed i}
\def\slj{\slashed j}
\def\slk{\slashed k}
\def\sll{\slashed l}
\def\slm{\slashed m}
\def\sln{\slashed n}
\def\slo{\slashed o}
\def\slp{\slashed p}
\def\slq{\slashed q}
\def\slr{\slashed r}
\def\sls{\slashed s}
\def\slt{\slashed t}
\def\slu{\slashed u}
\def\slv{\slashed v}
\def\slw{\slashed w}
\def\slx{\slashed x}
\def\sly{\slashed y}
\def\slz{\slashed z}
\def\slE{\slashed E}

\title{
\vskip 10mm
\large\bf
Searching for the Origin of CP violation in Cabibbo Suppressed D-meson 
Decays.
}

\author{David Atwood}
\affiliation{Dept. of Physics and Astronomy, Iowa State University, 
Ames,
IA 50011}
\author{Amarjit Soni}
\affiliation{ Theory Group, Brookhaven National Laboratory, Upton, NY
11973}

\begin{abstract}

The recent evidence of relatively large direct CP violation in $D^0$ 
decay at LHCb suggests that CP studies in the $D$ system may become an 
important new avenue for understanding CP just as studies in the B 
system have proven to be. The current level of CP violation could be 
consistent with the Standard Model or, perhaps, contain evidence of new 
physics. A clean Standard Model prediction of the CP violation in these 
decays would, of course, be important in understanding these results but 
hadronic uncertainties makes such a prediction difficult. In this paper, 
we make several suggestions to try seek the role of new physics. We 
propose that the hadronic enhancement needed to attribute the observed CP 
violation in D to two pseudoscalar modes may not operate for inclusive 
final states where it is likely that we will see asymmetries at the 
quark level expectation provided the source is the Standard Model. A 
simple way to implement this is to search for CP asymmetries in final 
states containing K and $\bar K$ but where the sum of their energies is 
less than the energy of the parent D. This is meant to ensure that the 
event belongs to an inclusive and not an exclusive sample. We also 
propose that CP asymmetries may be enhanced in modes where the tree is 
color suppressed. In particular, the final state $\rho^0\rho^0$ is of 
special interest because it consists of charged pions only and, in 
addition, it can have C-even P-odd triple product correlations; 
similarly $D_s \to \rho^0 K^+$ and $\rho^0 K^{* +}$ also appear 
interesting.  We also emphasize the use of CPT constraints leading to 
interesting correlations. We then consider how isospin symmetry can 
provide observables which are sensitive to certain classes of new 
physics and are small in the Standard Model. In particular, we discuss 
using isospin analysis in the decays $D\to\pi\pi$, $\rho\pi$ and 
$\rho\rho$ as well as in $D_s\to K^*\pi$.  We also consider how such 
analysis may eventually be supplemented by information about the weak 
phases in $D^0$ decay. In order to obtain this information 
experimentally, we consider various methods for preparing an initial 
state which is a quantum mechanical mixture of $D^0$ and $\overline 
D^0$. This may be done through the use of natural $D^0/\overline D^0$ 
oscillations; observing $D^0$ mesons which arise from $B_d$ or $B_s$ 
mesons which themselves are oscillating or from quantum correlations in 
$D^0$ pairs which arise from either $\psi^{\prime\prime}$ decay or 
B-meson decay. Observing CP violation in the magnitudes of decay 
amplitudes should be within the capability of experiments in the near 
future, however, obtaining the weak phases through the methods we discuss 
will likely require future generations of machines due to the large 
statistics 
that are likely to be needed.

\end{abstract}

\pacs{11.30.Er, 12.60.Cn, 13.25.Hw, 13.40.Hq}

\maketitle

\section{introduction}

Recent results~\cite{Aaij:2011in} from the LHCb provide evidence for CP 
violation in D-meson decays, in particular, 
$A_{CP}(K^+K^-)-A_{CP}(\pi^+\pi^-) =-0.82\pm 0.21{\rm(stat)} \pm 
0.11{\rm (syst)}\%$ giving a 3.5 sigma signal of CP violation.  CDF has 
reported ~\cite{arXiv:1111.5023} the modes separately obtaining 
$A_{CP}(K^+K^-)=-0.24\pm 0.22\pm0.09 \%$ and 
$A_{CP}(\pi^+\pi^-)=+0.22\pm 0.24 \pm 0.11 \%$. CDF has also 
directly measured the difference previously 
measured by LHCb and obtained\cite{newCDF} 
$A_{CP}(K^+K^-)-A_{CP}(\pi^+\pi^-) =-0.62\pm 0.21{\rm(stat)} \pm 
0.10{\rm (syst)}\%$.
A similar result of 
$A_{CP}(K^+K^-)-A_{CP}(\pi^+\pi^-) =-0.87\pm 0.41{\rm(stat)} \pm 
0.06{\rm (syst)}\%$ was also recently reported by BELLE at 
ICHEP2012~\cite{nakao.ichep.2012}.  Belle ~\cite{nakao.ichep.2012} also 
gave $A_{CP}(K^+K^-)=-0.32\pm 0.21\pm 0.09 \%$ and 
$A_{CP}(\pi^+\pi^-)=+0.55\pm 0.36 \pm 0.09 \%$.  In the LHCb result, the 
cancellation of experimental uncertainties between the two modes plays 
an important role in the extraction of a significant signal for the 
difference in the two asymmetries.  These measurements dominate the 
world average for the difference given by the HFAG group\cite{hfag}: 
$\Delta A_{CP} \equiv 
A_{CP}(K^+K^-)-A_{CP}(\pi^+\pi^-) =-0.678\pm 0.147\%$. The Belle 
results are particularly significant for individual modes, because the 
leptonic environment allows better detection of these two final states, 
so super KEK~\cite{Hashimoto:2004sm,Browder:2008em} and the Super-B 
Factory~\cite{Bona:2007qt} should be able to produce more precise 
results in the future especially for individual modes.

The LHCb result for the difference in the asymmetries appears to be 
large compared to the Standard Model(SM)~\cite{Bediaga:2012py} based on 
early expectations as we will discuss below. The weak phase arises in 
the SM from the CKM matrix. The relevant combination of CKM elements 
which gives the weak phase between the tree and penguin graphs is

\begin{eqnarray}
\left | \theta_W \right |
\approx
5.6\times 10^{-4}
\label{zeroth_estimate}
\end{eqnarray}

\noindent where we have expanded this in terms of the the Wolfenstein 
parametrization\cite{wolf_matrix}. As discussed in 
Section~\ref{cptsection} we expect the CP asymmetry on the {\it quark 
level} to be roughly of this size. For specific final states, hadronic 
effects will alter this expectation appreciably especially since charm 
quark is so light. Thus the central value for 
$A_{CP}(K^+K^-)-A_{CP}(\pi^+\pi^-)$ may not be inconsistent with the 
SM~\cite{ Feldmann:2012js, Franco:2012ck, Bhattacharya:2012ah, 33a,33b, MJ1211} 
but this is far from a proof that the SM is fully adequate and therefore 
the role of new physics cannot yet be ruled out~\cite{ Feldmann:2012js, 
Franco:2012ck, Li:2012cf, Cheng:2012xb}. There is, in fact, the 
intriguing possibility that various models for new physics (NP) are also 
able to contribute significantly to direct CP violation in 
D-decays~\cite{Agashe:2004cp, 
Nandi:2010zx,Buras:2010nd,BarShalom:2011zj, 
Rozanov:2011gj,Feldmann:2012js,Hiller:2012wf,LVecchi11, Delaunay:2012cz, 
Giudice:2012qq,Altmannshofer:2012ur,Mannel:2012hb,Chen:2012usa,Cheng:2012xb}.  
For this reason and others, it is important to devise observables which 
can distinguish between SM and NP origins for this CP violation.

Broadly speaking, for singly Cabibbo suppressed (SCS) decays it is 
useful to divide potential models of NP into two categories. The first 
is ``penguin like'' where the effective Hamiltonian is strictly $\Delta 
I=\frac12$. Generally this kind of contribution will result only when 
the NP contributes through a gluonic penguin. The second is ``tree 
like'' where the effective Hamiltonian contains a $\Delta I=\frac32$ 
component. This includes models where there are extra massive scalar or 
vector bosons which enter at tree level as well as photon, Z or W 
penguin topologies. In principle, electroweak penguins could contribute 
in this way but in the SM such contributions are negligible.

Depending on whether the new physics is tree like or penguin like, 
different kinds of studies may help identify the underlying mechanism. 
In this paper we consider strategies which would be helpful in both 
cases.

In section~\ref{cptsection} the requirements of CPT symmetry motivates 
us to devise a general test for SM versus NP. If the large observed CP 
asymmetry is due to SM alone, then we would expect CP violation in a 
more inclusive final state to converge to the SM quark level expectation 
given in eqn.~\ref{zeroth_estimate}. Using the number of kaons as a 
surrogate for the number of s-quarks, we suggest that CP violation in 
inclusive $K\overline K+X$ final states would provide a good test of 
this idea.  We also suggest that hadronic matrix element enhancements
 mostly occur in only exclusive two-body modes (especially pseudoscalars).
 With that in mind a simple method is suggested to experimentally
identify inclusive events.

In section~\ref{candidates} we focus on finding additional two body 
decay modes which are likely to also show large CP asymmetries in the 
case of penguin-like NP. The key point here is to go after color - 
suppressed tree modes.  In selecting potentially useful modes, we also 
consider which final states are more easily detected because they appear 
in the final state as charged particles only. Using these criteria modes 
of particular interest include $D^0\to\rho^0\rho^0$, 
$D_s\to\rho^0K^{(*)+}$ and $D^0\to K^{(*)0}\overline K^{(*)0}$. In 
addition, the final states $\eta\eta$, $\eta\eta^\prime$ and $\eta\phi$ 
may be interesting because the ``s-quark rich'' nature of the final states 
even though they contain neutrals in the final state; these could be of 
special interest to upcoming SuperB Factories. We briefly discuss 
radiative final states which are considered in~\cite{Isidori:2012yx}. 
These modes may show CP violation in a large class of penguin NP models, 
however through CPT arguments we show that $A_{CP}$ is likely to be 
suppressed.

In all of the above, CP violation is observed through $A_{CP}$, a 
difference in decay rate between $D$ and $\overline D$ to a given final 
state. For this form of CP violation a strong phase is also required. It 
would be very useful to also be able to measure the weak phase directly, 
independently of the strong phase. In section~\ref{phaseSection} we 
propose various methods to measure this phase. Three methods are 
considered: (1) using $D^0$ oscillation just as oscillation in $B$ 
mesons are used to measure the weak phase in $B^0\to \psi K_s$; (2) 
using oscillation in B-mesons where $B\to \overline D^0 \rho^0$ (or 
similar final states) and (3) using correlations in $D^0\overline D^0$ 
pairs. If we assume that the phase between $D^0$ and $\overline D^0$ 
decay to a given final state is of the same magnitude as the observed 
value of $A_{CP}$ in $D\to \pi\pi$ and $D\to KK$ then the statistics 
required is $\sim 10^{11}$ mesons for the methods using oscillation in 
D-mesons, oscillation in D-mesons and correlations in D-pairs 
originating from B-meson decay. In the case where the D-pairs arise in a 
$\psi^{\prime\prime}$ factory, then $\sim 10^9$ D-mesons are required. 
More realistically, weak phases an order of magnitude larger than the 
currently observed $A_{CP}$ may be observable in the foreseeable future. 
This would indicate a situation where the strong phase is small and yet 
there is a large NP weak phase.

In section~\ref{isospin} we consider tests for NP based on isospin which 
would apply to tree-like NP models. In some cases isospin can be used to 
isolate CP violation in the $\Delta I=\frac32$ channel. Since the SM 
predicts no CP violation in this channel, such a signal would indicate 
the presence of NP.

Tests of this form where the magnitude of a $D$ amplitude is compared to 
that of a $\overline D$ decay amplitude are proposed for $D\to \pi\pi$, 
$D\to \rho \pi$ , $D_s\to \pi K^*$ and $D\to \rho\rho$. In the case of 
$D\to\rho\pi$ there are two separate amplitudes which can be used, one 
derived only from the $D^0\to \pi^+\pi^-\pi^0$ overall reaction and one 
which also includes input from $D^+\to \rho^+\pi^0$; $\rho^0\pi^+$. In 
the case of $D\to\rho\rho$ each polarization can provide a separate 
test.

For the final states $\pi\pi$, $\rho\rho$ and $\rho\pi$, we can also 
combine this analysis with weak phase determination in 
section~\ref{phaseSection} then additional tests are possible where the 
phase of a $\Delta I=\frac32$ amplitude is compared to its conjugate. 
Again such a phase would be indicative of tree-like NP.

In Section~\ref{numerics} we discuss the statistical requirements for 
testing the SM, particularly for the determination of weak phases. In 
Section~\ref{conclusion} we give our summary and conclusion.

\section{CPT and Flavor Symmetry Considerations}
\label{cptsection}

\subsection{Rough Estimate for Quark Level Expectations}

The effective Hamiltonian for SCS charm decay can be written:

\begin{eqnarray}
H_{eff} &=&
\frac{G_F}{\sqrt{2}}
\bigg\{
\frac12 (\lambda_s-\lambda_d)\sum_{i=1,2}C_i(Q_i^s-Q_i^d)
-
\lambda_b
\left ( 
\sum_{i=1,2}\frac12 C_i(Q_i^s+Q_i^d)
+
\sum_{i=3,6}C_iQ_i
\right )
\bigg\}
+h.c.
\end{eqnarray}

\noindent
where $\lambda_q=V_{cq}V^*_{uq}$ (note that 
$\lambda_d+\lambda_s+\lambda_b=0$ by CKM unitarity). The operators are  

\begin{eqnarray}
Q_1^q=(\overline q u)_{V-A}(\overline c q)_{V-A}
\ \ \ \ \ 
Q_2^q=(\overline q_\alpha u_\beta)_{V-A}(\overline c_\beta 
q_\alpha)_{V-A}
\nonumber\\
Q_3=\sum_{q=u,d,s}
(\overline c u)_{V-A}(\overline q q)_{V-A}
\ \ \ \ \ 
Q_4=\sum_{q=u,d,s}
(\overline c_\alpha u_\beta)_{V-A}(\overline q_\beta 
q_\alpha)_{V-A}
\nonumber\\
Q_5=\sum_{q=u,d,s}
(\overline c u)_{V-A}(\overline q q)_{V+A}
\ \ \ \ \ 
Q_6=\sum_{q=u,d,s}
(\overline c_\alpha u_\beta)_{V+A}(\overline q_\beta 
q_\alpha)_{V-A}
\nonumber\\
\end{eqnarray}

The first term proportional to $\lambda_s-\lambda_d$ is the tree 
contribution and the term proportional to $\lambda_b$ is the penguin 
contribution. If we assume that there is a strong phase difference 
between the tree and penguin of $\phi_{strong}$ then we can write the CP 
asymmetry in the quark level process $c\to d\overline d u$ 
as~\cite{Bander:1979px}:

\begin{eqnarray}
\left | A_{CP}(c\to d\overline du) \right | 
&=& 
\left | Im\left (  \frac {2\lambda_b}{\lambda_s-\lambda_d }   \right )
\right |
R\sin\phi_{strong}
\approx
\left |
Im\left(
\frac
{V_{ub}V_{cb}^*}
{V_{us}V_{cs}^*}
\right)
\right |
R\sin\phi_{strong}
\nonumber\\
&&\approx
A^2\lambda^4\eta
R\sin\phi_{strong}
\approx
6.4\times 10^{-4}
\sin\phi_{strong}
\label{first_estimate}
\end{eqnarray}

\noindent Here $R$ is a number of order 1 which depends on the Wilson 
coefficients. If we neglect the mass of the s-quark and hadronization 
effects and use the one loop evolution of the Wilson coefficients given 
in \cite{Golden:1989qx,Abbott:1979fw} then numerically $R \approx 1.2$. 
The resultant asymmetry at the quark level is thus expected to be about 
the same as given in Eqn.~\ref{zeroth_estimate} if the strong phase is 
near maximal. As discussed below CPT implies $\Delta\Gamma(c\to 
d\overline du)=- \Delta\Gamma(c\to s\overline su)$ where for a given 
decay $A\to B$, $\Delta\Gamma(A\to B)=\Gamma(A\to B)- \Gamma(\overline 
A\to \overline B)$

This quark level result need not be the same as the CP asymmetry in any 
given exclusive hadronic final state. Due to the significant hadronic 
uncertainties in the formation of specific final states, in order to 
characterize the CP violation in D-meson decay it is useful to consider 
symmetries which may be partially respected by strong interactions. The 
most obvious such symmetry is $SU(3)_{flavor}$ which, unfortunately is 
badly broken~\cite{Cheng:2012xb}. We will discuss the use of the isospin 
subgroup of $SU(3)$ in section~\ref{isospin}. For the current 
experimental results, a more handy subgroup of $SU(3)$ to consider is 
U-spin since it directly relates the $K^+K^-$ and $\pi^+\pi^-$ final 
states. If this symmetry were strictly observed then 
$Br(D^0\to\pi^+\pi^-)=Br(D^0\to K^+K^-)$ and 
$A_{CP}(D^0\to\pi^+\pi^-)=-A_{CP}(D^0\to K^+K^-)$. Since 
$Br(D^0\to\pi^+\pi^-)=(1.397\pm 0.026)\times 10^{-3}$ while $Br(D^0\to 
K^+K^-)=(3.94\pm 0.07)\times 10^{-3}$, more than a factor of 2 
discrepancy, it is clear that U-spin is badly broken. Due to the large 
experimental error in the individual CP asymmetries, no firm conclusion 
with respect to U-spin applied to CP violation can be drawn though the 
central values tend to show opposite sign.

\subsection{CPT Relations}

Another symmetry which must, of course, be respected, is CPT which 
implies that the width of the $D$ and $\overline D$ mesons must be the 
same. This means that when summed over all final states of D-meson 
decay, the partial rate differences must vanish:

\begin{eqnarray}
\sum_X \Delta\Gamma(X)=0
\label{pra:general}
\end{eqnarray}

In detail, this means ~\cite{Gerard:1988jj,Atwood:2000tu} that 
$\Delta\Gamma$ must be exchanged between the various final states. This 
exchange is caused by rescattering between at least two final states 
(say $X_1$ and $X_2$) with different strong and weak phases. If $X_2$ 
rescattering into $X_1$ provides a strong phase for $X_1$ it will give 
rise to a contribution to $\Delta\Gamma(X_1)$. This partial rate 
asymmetry will be exactly canceled by the contribution to 
$\Delta\Gamma(X_2)$ proportional to the strong phase produced when $X_1$ 
rescatters into $X_2$.

At the quark level, the SM maintains CPT in SCS decays by an exchange of 
$\Delta \Gamma$ between $c\to u d\overline d$ and $c\to u s\overline s$, 
hence

\begin{eqnarray}
\Delta\Gamma(c\to d\overline du)
=
-\Delta\Gamma(c\to s\overline su).
\label{CPTquark}
\end{eqnarray}

\begin{figure}
{
\includegraphics[angle=0,
width=0.6\textwidth]{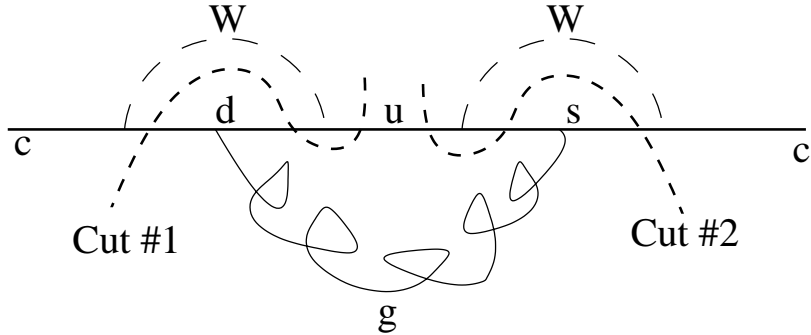}
}
\caption{
The unitarity graph showing the CPT identity Eqn.~\ref{CPTquark} for the 
quark level SCS charm decay. Cut \#1 indicated in the figure shows the 
case where the decay is $c\to d\overline du$ with a $s\overline su$ 
intermediate state providing the strong phase. Conversely, cut \#2 
indicated in the figure shows the case where the decay is $c\to 
s\overline su$ with a $d\overline du$ intermediate state providing the 
strong phase. The interfering tree graphs are not shown but are implied
}
\label{dcp_CPT1}
\end{figure}

\noindent The ``double penguin'' unitarity graph in 
Fig.~\ref{dcp_CPT1} shows how this compensation arises where the two cuts 
indicate the two final states. Thus, cut \#1 gives a final state with $d 
\overline du$ where one of the amplitudes has an internal loop with an $s 
\overline su$ final state. The magnitude of this graph is the same as 
that given by cut \#2 giving a $s \overline su$ final state with an 
intermediate $d \overline du$ but the sign is opposite due to the 
internal loop being on the left side in this case.

\begin{figure}
{
\includegraphics[angle=0,
width=1.2\textwidth]{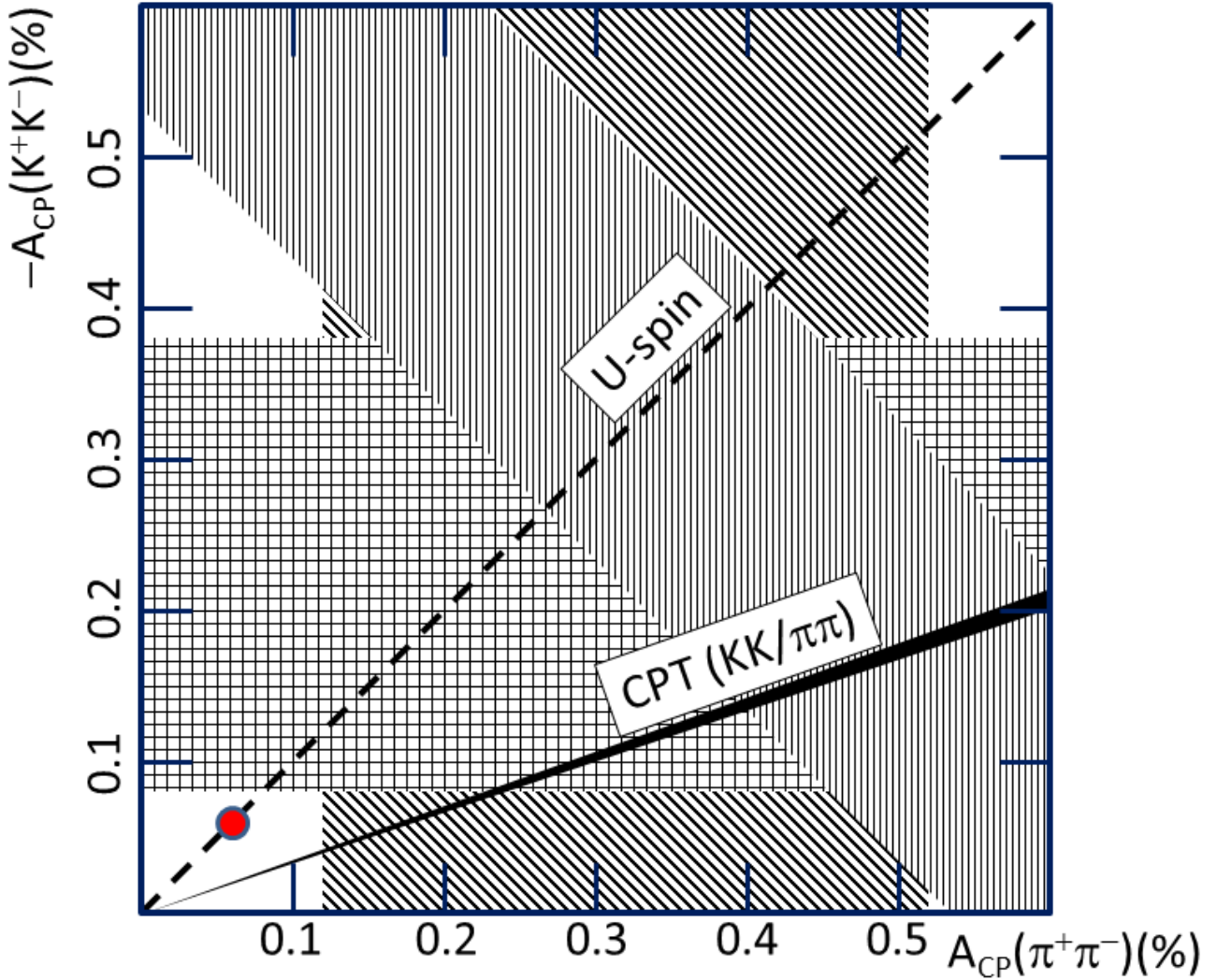}
}
\caption{
The current experimental results for $A_{CP}(\pi^+\pi^-)$ and 
$A_{CP}(K^+K^-)$. The vertically hatched band shows the 1-$\sigma$ region 
for the measurement of $A_{CP}(\pi^+\pi^-)-A_{CP}(K^+K^-)$; the diagonally 
hatched band shows the 1-$\sigma$ region for the measurement of 
$A_{CP}(\pi^+\pi^-)$; the square hatched band shows the 1-$\sigma$ region 
for the measurement of $A_{CP}(K^+ K^-)$. The dashed line indicates the 
U-spin prediction that $A_{CP}(\pi^+\pi^-)+A_{CP}(K^+K^-)=0$. The black 
wedge indicates the result where CPT is maintained between $K^+K^-$ and 
$\pi^+\pi^-$, i.e. $\Delta\Gamma(K^+K^-)+\Delta\Gamma(\pi^+\pi^-)=0$. The 
dot towards the lower right is the quark level expectation from 
Eqn.~\ref{first_estimate} with maximal strong phase if you make the 
naive assumption that $A_{CP}(K^+K^-)=A_{CP}(s\overline su)$ and 
$A_{CP}(\pi^+\pi^-)=A_{CP}(d\overline du)$
}
\label{Acp_graph}
\end{figure}

To draw conclusions concerning specific groups of decay modes, it is 
useful to break down Eqn.~\ref{CPTquark} according to quantum numbers 
conserved by strong interactions, since the exchange of $\Delta\Gamma$ 
between final states can only occur between states which can rescatter 
into each other. Such rescattering is via strong interactions so the 
general statement in Eqn.~\ref{pra:general} can be refined to:

\begin{eqnarray}
\Delta\Gamma(
{P,C,I,G,S}
)=0
\label{pra:strongConstrained}
\end{eqnarray}

\noindent
where
${P,C,I,G,S}$
are the quantum numbers, parity, charge conjugation, isospin, G-parity 
and strangeness respectively.

In applying this to SCS modes, where $S=0$ in the final state, we can 
further classify final states according to the number, $N_K$, of kaons 
and anti-kaons they contain. Notice that in general $N_K\in\{0,1,2,3\}$ 
because $M_D<4M_K$ and for S=0 in particular, $N_K\in\{0,2\}$ therefore

\begin{eqnarray}
\Delta\Gamma(PCIG,S=0,N_K=0)
=
-\Delta\Gamma(PCIG,S=0,N_K=2)
\label{K:balance}
\end{eqnarray}

\noindent The left and right sides of these equations should represent 
$d\overline du$ and $s\overline su$ quark content respectively since it 
is expected that $c\to u d\overline d$ couples dominantly to $N_K=0$ 
while $c\to u s\overline s$ couples dominantly to $N_K=2$ so 
$\Delta\Gamma(c\to u d\overline d)\sim \Delta\Gamma(N_K=0)$ and 
$\Delta\Gamma(c\to u s\overline s)\sim \Delta\Gamma(N_K=2)$.

Implementing Eqn.~\ref{K:balance} directly for each combination of 
quantum numbers is difficult since many of the D decay modes contain 
multiple charged pions and so the quantum numbers may be difficult to 
determine on an event by event basis.

What may be a more practical an experimental test of CP violation is to 
look for CP violation in the inclusive case summed over $P,C,I,G$. In 
this case CPT implies:

\begin{eqnarray}
\hat\Delta\Gamma(S=0,N_ K=0) + \Delta\Gamma( \pi + \pi)
=
-\hat\Delta\Gamma(S=0,N_K=2) - \Delta\Gamma( K ~ K )
\label{K:balance:all}
\end{eqnarray}

\noindent 
where $\hat\Delta$ means that two body pseudoscalars are not included as we explain
more below.
CP asymmetry in both $\hat\Delta\Gamma(S=0,N_K=0)$ and 
$\hat\Delta\Gamma(S=0,N_K=2)$ should approximate the quark level CP asymmetry 
in $d \overline du$ and $s \overline su$ respectively.

If CP violation in the $PP$ final states is due to the SM then there 
must be some hadronic enhancement for those exclusive final states such 
as $\pi^+\pi^-$, $K^+K^-$, which we would not expect to be present in 
the inclusive Eqn.~\ref{K:balance:all}. Recall also that for exclusive 
two pseudosaclar modes, in particular, there are well known reasons to 
expect large QCD corrections, {\it e. g.} chiral enhancements. It is 
quite unlikely that inclusive modes will receive such largish 
enhancements. Thus, the inclusive CP asymmetry should be smaller, $\sim 
6\times 10^{-4}$. On the other hand, if the largeness of the CP 
asymmetry in the PP final states is due to NP, then one would expect the 
inclusive asymmetry to be roughly of the same order as the exclusive. 
The larger statistics of the inclusive state may provide more accurate 
results for this channel leading to an important indication of the 
nature of the observed CP violation.

In practice, observing a quantity like $\hat\Delta\Gamma(S=0,N_K=2)$ is 
subject to the problem that it is not possible to catch every final 
state. Thus it is useful to rephrase the relation 
$\hat\Delta\Gamma(S=0,N_K=2)\sim \hat\Delta\Gamma(c\to s \overline su)$ 
as

\begin{eqnarray}
lim_{\chi\to I}
\hat\Delta\Gamma(S=0,N_K=2;\chi)
\sim \hat\Delta\Gamma(c\to s \overline su)
\end{eqnarray}

\noindent
where $\chi$ is some CP invariant acceptance cut on the final states
and $I$ represents the cut where all events are accepted. In any case 
$\hat\Delta\Gamma(S=0,N_K=2;\chi)$ is a CP violating quantity.

Actually, given that in the sample of inclusive ($N_K = 2$) final states 
we do not want to include the exclusive $K\overline K$ mode, a simple 
working definition of inclusive is all those final states in which the 
sum of $K$ and $\overline K$ energies is less than the energy of the parent 
$D$.

Eqn.~\ref{K:balance} can be broken down further with approximate 
symmetries allowing us to gain some understanding of the pattern of CP 
violation in exclusive decay modes. For instance, if U-spin were a good 
symmetry then Eqn.~\ref{K:balance} could be broken down into groups of 
final states related by this symmetry. In particular it would follow 
that $\Delta\Gamma(\pi^+\pi^-)=-\Delta\Gamma(K^+K^-)$. Of course U-spin 
is broken but in~\cite{Feldmann:2012js} figure 2 they fit the 
experimental data including U-spin breaking allowed within the SM. This 
fit favors a solution where $\left | A_{CP}(\pi^+\pi^-) / 
A_{CP}(K^+K^-)\right | >1$ although the statistics are not yet good 
enough to draw any firm conclusion. Analogously, in~\cite{33b} (see eq 
34), the above ratio of asymmetries is predicted to be $\approx 1.8$. 
This suggests a pattern where the partial rate asymmetry exchange is 
mostly between these two final states, in which case we would expect $ 
A_{CP}(\pi^+\pi^-) / A_{CP}(K^+K^-) \approx - Br(K^+K^-) / 
BR(\pi^+\pi^-) \approx -2.82\pm 0.14$.

In Figure~\ref{Acp_graph} the current weighted average results for 
$A_{CP}(\pi^+\pi^-)$ and $A_{CP}(K^+K^-)$ are shown as 1$\sigma$ bands 
on a $A_{CP}(\pi^+\pi^-)$ versus $A_{CP}(K^+K^-)$ plot (indicated by the 
diagonally hatched and square hatched regions respectively) as well as 
the world average for the $A_{CP}(K^+K^-)-A_{CP}(\pi^+\pi^-)$ result 
(vertically hatched band). The dashed line indicates the U-spin result 
$A_{CP}(K^+K^-)= - A_{CP}(\pi^+\pi^-)$ while the black wedge indicates 
the result where we assume that CPT is maintained within PP final states 
by exchange between $K^+K^-$ and $\pi^+\pi^-$, i.e. that 
$\Delta\Gamma(K^+K^-)+\Delta\Gamma(\pi^+\pi^-)=0$. For comparison the 
circle in the lower left corner indicates the naive expectation where we 
assume that the meson asymmetry is the same as the quark level 
asymmetry. In particular we suppose here that the quark level asymmetry 
is given by Eqn.~\ref{first_estimate} with maximal strong phase and that 
$A_{CP}(K^+K^-)=A_{CP}(s\overline su)$ and 
$A_{CP}(\pi^+\pi^-)=A_{CP}(d\overline du)$.

In order to facilitate distinguishing SM from NP contributions, in 
Section (\ref{isospin}) we will discuss relations which rely on isospin 
only which, unlike the more general $SU(3)$, should be good to the level 
of a few percent.

\section{Candidates for Enhanced CP violation for Penguin-Like New 
Physics}
\label{candidates}

Suppose that CP violation is the result of a large amplitude $A$ 
interfering with a smaller amplitude $a$. If we normalize the amplitudes 
in units of square root of branching ratio, then $Br(D\to f)\approx \left 
|A\right |^2$ while $A_{CP}(f)\propto a/A$. If we want to observe the CP 
violation with a significance of $N_\sigma$, the number of mesons 
required is $N=N_\sigma^2/(Br A_{CP}^2)$. In terms of the amplitudes 
then,

\begin{eqnarray}
N=N_\sigma^2/(Br A_{CP}^2)
\propto \frac{N_\sigma^2}{|A|^2 |a/A|^2   }
\propto \frac {N_\sigma^2}{|a|^2}
\end{eqnarray}

So that generally $N$ depends on $a$ but is independent of $A$ but a 
smaller value of $A$ does enhance $A_{CP}$; $N$ is not affected because 
this is at the expense of the  branching ratio. Going to a mode which has 
smaller branching ratio with higher asymmetry has the advantage of 
reducing the effects of systematic errors and other errors which are not 
statistical in nature, {\it all other things being equal}.

If we assume that the observed CP violation in $D^0\to \pi^+\pi^-$, 
$K^+K^-$ is due to penguin like NP it may be that larger signals of CP 
asymmetries will be present in similar decays where the SM tree 
contribution is suppressed. Following this rationale, in this section we 
focus on the cases where the SM tree is color suppressed.

Color suppression in two body final states is a pattern which is often 
born out in B-meson decays. For example in decays to charm mesons the 
color allowed $B^0\to \pi^+ D^-$ has a branching ratio of $(2.6\pm 
0.13)\times 10^{-3}$ while the analogous color suppressed mode $B^0\to 
\pi^0 \overline D^0$ has a branching ratio an order of magnitude smaller 
of $(2.61\pm 0.24)\times 10^{-4}$. A similar pattern obtains for related 
decays. In contrast, in the case of B-meson decay to two light 
pseudoscalar mesons, color suppression fails. Thus $Br(B^0\to 
\pi^+\pi^-)=(5.13\pm 0.24)\times 10^{-6}$ while $Br(B^0\to 
\pi^0\pi^0)=(1.62\pm 0.31)\times 10^{-6}$ but with PV and PP final 
states, for instance $Br(B^0\to \pi^+\rho^-+\pi^-\rho^+)=(2.3\pm 
0.23)\times 10^{-5}$ versus $Br(B^0\to \pi^0\rho^0)=(2.0\pm 0.5)\times 
10^{-6}$, likewise $Br(B^0\to \rho^+\rho^-)=(2.42\pm 0.31)\times 
10^{-5}$ versus $Br(B^0\to \rho^0\rho^0)=(7.3\pm 2.6)\times 10^{-7}$; 
color suppression of these modes seems to hold.

It is to be expected that color suppression is less effective in $D$ 
decays because of greater non-perturbative effects and increased meson 
rescattering at the charm mass scale. Indeed this appears to be the 
case, for example $Br(D^0\to K^-\pi^+)=3.89\%$ while $Br(D^0\to \bar 
K^0\pi^0)=2.44\%$ showing no color supression, likewise $Br(D^0\to 
K^{*-}\pi^+)=1.73\%$ while $Br(D^0\to \overline K^{*0}\pi^0)=2.28\%$ 
again showing no color suppression. Conversely the $K\rho$ channel does 
seem to show the effect since $Br(D^0\to K^{-}\rho^+)=10.8\%$ while 
$Br(D^0\to \overline K^{0}\rho^0)=1.32\%$.

In spite of the unreliable evidence that color suppression is 
universally operative in D-meson decays, used with care and caution, it 
may provide us with a guide in searching for other decay modes for 
future searches for enhanced CP violation. In particular, modes where 
the SM amplitude tends to be suppressed and a possible NP penguin 
amplitude may be enhanced may be of particular interest in NP searches. 
In table~\ref{mode_table} we list all the two body Cabibbo suppressed 
decay modes of D-mesons to ground state mesons. Here we use the notation 
$\pi^{(*)\pm}$ to indicate either $\pi^\pm$ or $\rho^\pm$ and 
$\pi^{(*)0}$ to indicate either $\pi^0$, $\rho^0$ or $\omega^0$. 
Likewise we use $\phi^{(*)}$ to indicate either $\phi$ or 
$\eta^{(\prime)}$ (or at least the $s\overline s$ component of the 
latter).

Decays of the form $D_s\to \pi^{(*)0}K^{(*)+}$, $D^+\to 
\pi^{(*)+}\phi^{(*)}$, $D^0\to K^{(*)0}\overline K^{(*)0}$ and $D^0\to 
\pi^{(*)0}\phi^{(*)}$, have the tree color suppressed in this way. Also 
modes of the form $D^0\to K^{(*)0}\overline K^{(*)0}$ have an additional 
suppression of the tree contribution due to the fact that the 
$d\overline d s\overline s$ final state quark content is not the same as 
produced by the tree graph. This is born out by the smallness of the 
branching ratios of $D^0\to K_sK_s$ and $D^0\to K^{*0}\overline K^{*0}$. 
Therefore, most promising from this list is $D_s\to \pi^{(*)0}K^{(*)+}$.

As an example of how color suppression can work in the realm of Cabibbo 
Allowed D-meson decays, consider such decays to two vector final states. 
The decay $D^0\to K^{*+}\rho^-$ has no suppression and its branching 
ratio is $10.8\pm0.7$\%. The related color suppressed decays $D^0\to 
\overline K^{*0}\rho^0$ and $D^0\to \overline K^{*0}\omega$ have 
branching ratios $1.58\pm 0.35\%$ and $1.1\pm 0.5\%$ respectively.

In the table, we also enumerate the cases where the mode cascades down 
to a final state which contains all charged particles (i.e. $\pi^\pm$ 
and $K^\pm$), and give their branching ratios from~\cite{pdb}, where 
known.  Final states with all charged states will generally be easier to 
detect, particularly at the LHCb. Based on these criteria, $D^0\to 
\rho^0\rho^0$, $D_s\to \rho^0 K^+$ and $D_s\to \rho^0 K^{*+}$ are 
perhaps the most favorable channels to find CP violation due to penguin 
like new physics.

From this point of view, the cases of $D^0\to \rho^0\rho^0$ is perhaps of 
particular interest to search for enhanced CP violation due to NP. An 
additional feature of this VV final state is the spin degree of freedom: 
there are three polarization states: transverse parallel ($A_\parallel$), 
transverse perpendicular ($A_\perp$) and longitudinal ($A_\ell$) . Each 
amplitude could have different CP violation. Already existing 
measurements ~\cite{pdb} of the polarization fractions show that 
the longitudinal mode dominates with a fraction of 67\% longitudinal. 
Further measurements of the angular distribution using the methods 
of~\cite{Chiang:1999qn} will allow the extraction of the phases between 
the polarization amplitudes.

As discussed in~\cite{Bigi:2011em,Gronau:2011cf} and 
in~\cite{Datta:2011qz} for the analogous B decays, a qualitatively 
different feature of VV final states is that there can be P-odd triple 
product observables. Such observables can lead to either CP-odd or 
CP-even correlations depending on the combination of $D$ and $\overline 
D$ decays. If the C-even combination is formed (adding the triple 
product of $D$ and $\overline D$) then the combination is CP-odd 
conversely the C-odd combination is CP-even. CPT then requires C-odd, CP 
violating amplitudes to be real whereas C-odd, CP violating amplitudes 
need a rescattering strong phase.

CP-odd observables of this form that are C-even can be formed from 
untagged samples of $D^0$ mesons. This is an advantage for $e^+e^-$ 
B-factories where the initial state is self conjugate so the $D^0$ 
samples obtained at such machines could be used directly (except for the 
small asymmetry in the D meson production mechanism between inclusive 
$B\to \overline D+X$ versus $\overline B\to D+X$ which will have to be 
determined from separate studies). In the case of LHCb this would not be 
true since the $pp$ initial state is not self conjugate so tagging will, 
in any case, be necessary.

In some tree like NP models, even if the SM tree is colored suppressed 
the NP contribution is not. This will tend to enhance the NP 
contribution to CP violation. In particular, if a $q\overline q$ pair is 
produced by a color neutral object (e.g. a $Z^\prime$ or higgs like 
boson) then the effective Hamiltonian will have a different color 
structure form the SM and so color suppression may not apply. In 
addition to the two families of modes mentioned above, $D_s\to 
\pi^{(*)0}K^{(*)+}$, and $D^0\to \pi^{(*)0}\pi^{(*)0}$, modes of the 
form $D^+\to \pi^{(*)+}\phi^{(*)}$ should have enhanced CP asymmetries 
in this scenario.

Treelike CP violation of this form should contribute to the $\Delta 
I=\frac32$ channel and so the isospin analysis in Section~\ref{isospin} 
should reveal this though with some unknown hadronization effects and 
thus may reveal a contradiction with the SM. In this scenario, the 
$\pi\pi$, $\rho\pi$ and $\rho\rho$ systems are particularly suited since 
there are enough charge distributions to allow isospin analysis and the 
final state with two neutral mesons has a color suppressed tree graph 
which may enhance NP CP violation. Of these cases, the $\rho^0\rho^0$ 
state also has the advantage that it leads to an observed final state 
with all charged mesons.

Another class of final states which may be of interest in some models of 
new physics are those which are rich in $s\overline s$, in particular if 
they contain $\phi$ $\eta$ and $\eta^\prime$. The only three such two 
body modes which are kinematically allowed are $D^0\to\phi\eta$, 
$D^0\to\eta^\prime \eta$ and $D^0\to\eta\eta$. Of course none of these 
modes leads to an all charged final state but the tree graph is color 
suppressed. There is some additional suppression since the quark content 
only couples to the $u\overline u$ part of the $\eta^{(\prime)}$ wave 
function which makes up only 20-30\% of these mesons.

Another manifestation of penguin-like NP which could lead to CP 
violating signals are radiative decays. At the quark level such decays 
would proceed through $c\to u \gamma$ which leads to modes like $D^0\to 
\rho^0\gamma$, $D^0\to \omega\gamma$, $D^+\to \rho^+\gamma$ and $D_s\to 
K^{*+}\gamma$. Other radiative D-meson decays which would not be 
expected to receive large contributions from short distance radiative 
penguins are $D^0\to K^{0*}\gamma$ ($Br=(3.28\pm 0.35 )\times 10^{-4}$), 
$D^0\to \phi\gamma$ ($Br=(2.70\pm 0.34 )\times 10^{-5}$). 
In~\cite{Isidori:2012yx} these kind of radiative decays are discussed in 
the context of new physics. According to their analysis, if the QCD 
dipole $c\to u$ transition operators:

\begin{eqnarray}
Q_8=g_s\frac{m_c}{4\pi^2}\overline u_L
\sigma_{\mu\nu}T^aG^{\mu\nu}_a c_R
\ \ \ \ 
Q_8^\prime=g_s\frac{m_c}{4\pi^2}\overline u_R
\sigma_{\mu\nu}T^aG^{\mu\nu}_a c_L
\end{eqnarray}

\noindent
then operator evolution from the NP scale to the charm scale would lead 
to comparable coefficients for the electromagnetic dipole $c\to u$ 
transition operators:

\begin{eqnarray}
Q_7=eQ_u \frac{m_c}{4\pi^2}\overline u_L
\sigma_{\mu\nu} F^{\mu\nu} c_R
\ \ \ \ 
Q_7^\prime=eQ_u \frac{m_c}{4\pi^2}\overline u_R
\sigma_{\mu\nu}  F^{\mu\nu} c_L
\end{eqnarray}

\noindent Even if the coefficient of $Q_7^{(\prime)}$ were much smaller 
than $Q_8$ at the high NP scale, the coefficients may be comparable at 
the charm scale. Assuming that the observed LHCb result is due to the 
effects of $Q_8^{(\prime)}$ they~\cite{Isidori:2012yx} suggest that the 
induced coefficient of $Q_7^{(\prime)}$ would generate $A_{CP}$ in $D\to 
\rho\gamma$ of $O(10\%)$ provided the strong phases involved were 
maximal (See however\cite{zwi}). 
Using vector dominance one can estimate that $Br(D\to 
\rho\gamma)\approx 10^{-5}$ perhaps making this mode a good test for new 
physics.

\begin{figure}
{
\includegraphics[angle=0,
width=0.6\textwidth]{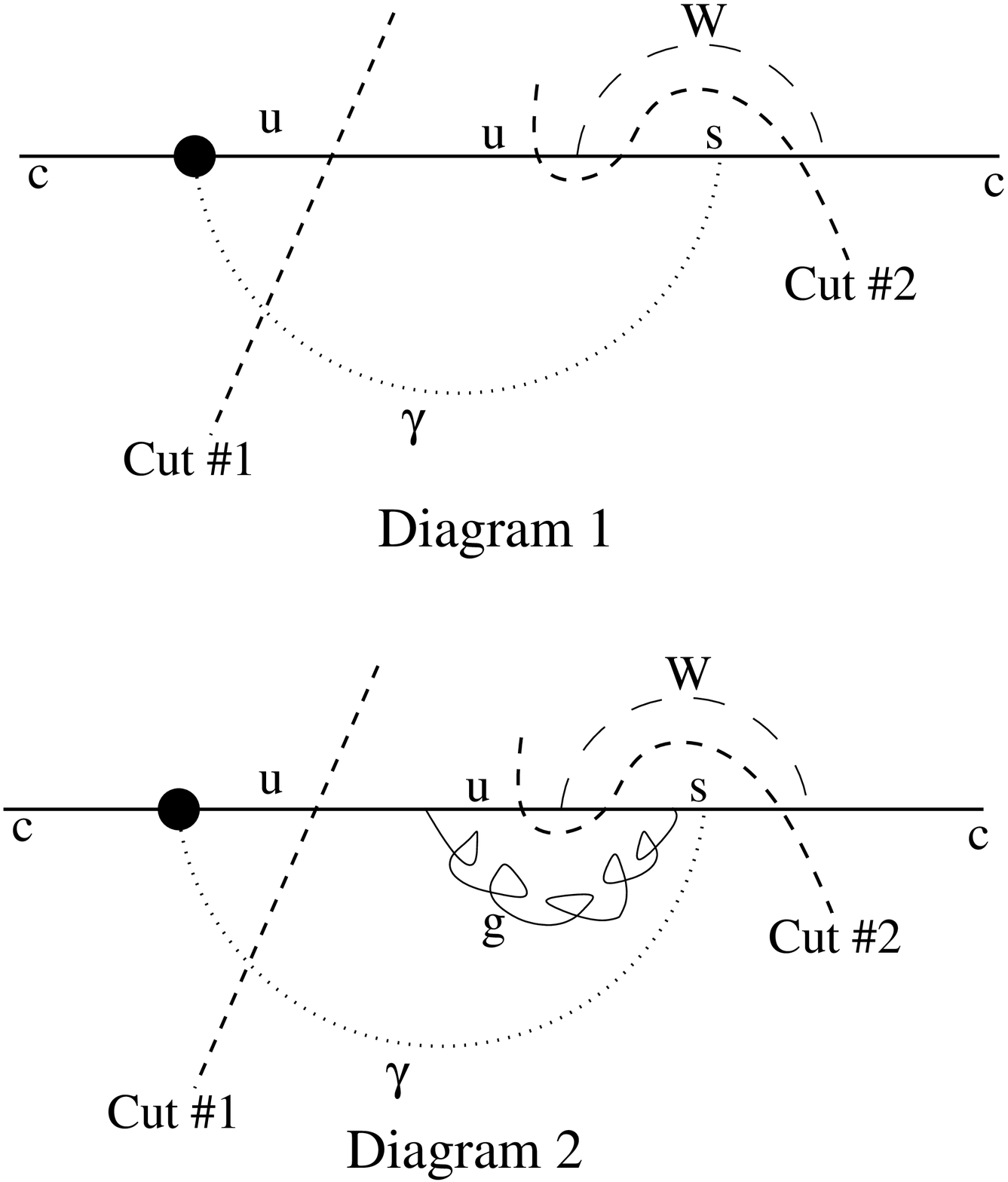}
}
\caption{
This unitarity graph illustrates CPT conservation for the quark level 
process $c\to u\gamma$ due to NP. Diagram 1 shows the lowest order 
interference between NP and SM where cut \#1 is for the $c\gamma$ final 
state and cut \#2 is for a $s\overline s u$ final state. Cut \#2 cannot 
be on shell. Diagram 2 shows an example of an order $\alpha_s$ 
correction to diagram 1 where in contrast cut \#2 can be on shell.
}
\label{dcp_CPT2} 
\end{figure}

It is, however, unlikely that the strong phase will be $O(1)$ because 
the kinematics of the rescattering forces there to be an $\alpha_s$ 
correction to satisfy CPT. To see this consider the unitarity diagrams 
in Figure~\ref{dcp_CPT2} for the quark level process $c\to u\gamma$. 
Diagram 1 in this figure shows the interference of a NP penguin with the 
SM penguin (having an internal s quark) at lowest order. Note that while 
cut \#1 corresponds to the $u\gamma$ final state, cut \#2 does not 
correspond to an on shell state since the $s\overline s$ pair must 
rescatter into a single photon. There is therefore no strong phase and 
so there cannot be a CP asymmetry from this diagram. Diagram 2 shows an 
order $\alpha_s$ correction to diagram where there can be a strong phase 
since now cut \#2 corresponds to an on shell state. This diagram 
therefore can give rise to a CP asymmetry but that asymmetry will be 
suppressed by $\alpha_s$ since there is an extra loop. Model calculations carried out 
in\cite{zwi} appear to bear out this situation.

\begin{table}[h]
\begin{tabular}{|c|c|c|c|c|}\hline
Decay & Suppressed    & Charged    & Favored & Total   \\ 
          &  Tree             &  Final State 
&              &  BR ($10^{-3}$)  \\ 
\hline
$D_s\to\pi^{(*)0} K^{(*)+}$  &X&  
$[\rho^0\to\pi^+\pi^-]K^+$ &X &   $2.7\pm 0.05$  \\
&&         $[\rho^0\to\pi^+\pi^-][K^{*+}\to \pi^+[K_s\to\pi^+\pi^-]]$ 
&X
& --- \\
\hline
$D_s\to\phi^{(*)} K^{(*)+}$  & &  $[\phi\to K^+K^-]K^+$&& $<0.3$   \\
&&   
$[\phi\to K^+K^-][K^{*+}\to \pi^+[K_s\to\pi^+\pi^-]]$&& --- \\
\hline
$D^+\to\pi^{(*)+} \phi^{(*)}$  
&X& $\pi^+[\phi\to K^+K^-]$ & X& $2.65\pm 0.08 $\\
\hline
$D^+\to K^{(*)+} \overline K^{(*)0}$  
      & & $K^+[K_s\to\pi^+\pi^-] $& & $1.98\pm 0.13$ \\
      & & $K^+[\overline K^{*0}\to K^+\pi^-] $& & $2.45^{.09}_{.14}$  \\
      & & 
$[K^{*+}\to\pi^+[K_s\to\pi^+\pi^-]][K_s\to\pi^+\pi^-] $& & $5.7\pm 2.3$\\
      & & $[K^{*+}\to\pi^+[K_s\to\pi^+\pi^-]][\overline K^{*0}\to 
K^+\pi^-] $ & & --- \\
\hline
$D^+\to \pi^{(*)+}  \pi^{(*)0}$  
& &  $\pi^+[\rho^0\to\pi^+\pi^-]$& & $0.81\pm 0.15$\\
\hline
$D^0\to K^{(*)0}\overline K^{(*)0}$ 
    &XX&   $[K_s\to\pi^+\pi^-][K_s\to\pi^+\pi^-]$ &$X$ & $0.085\pm 
0.014$ \\ 
    &  & $[K^{*0}\to K^+\pi^-][K_s\to\pi^+\pi^-]$ &$X$ & $<0.2$\\
    &  & $[\overline K^{*0}\to K^-\pi^+][K_s\to\pi^+\pi^-]$ &$X$ & 
$<0.35$\\
    & &  $[K^{*0}\to K^+\pi^-][\overline K^{*0}\to\pi^+K^-]$ &$X$ & 
$.07\pm 0.05$\\
\hline
$D^0\to \pi^{(*)0}\pi^{(*)0}$ 
	&X&  $[\rho^0\to\pi^+\pi^-][\rho^0\to\pi^+\pi^-]$ &X & 
$1.82\pm 0.10$ \\
\hline
$D^0\to \pi^{(*)+} \pi^{(*)-}$ 
    & &  $\pi^+\pi^-$ & & $1.400\pm .026$\\
\hline
$D^0\to\phi^{(*)}\pi^{(*)0}$
& X  & $D^0\to\phi\rho^0$&X  &  $1.40\pm 0.12$ \\
\hline
$D^0\to K^{(*)+} K^{(*)-}$ 
    & &  $K^+ K^-$ &  & $3.96\pm.08$\\
    & &  $[K^{*+}\to \pi^+[K_s\to\pi^+\pi^-]] K^-$& & $2.19\pm 0.1$ \\
    & &  $K^+[K^{*-}\to \pi^-[K_s\to\pi^+\pi^-]]$ & & $0.78\pm 0.06$\\
&& $[K^{*+}\to \pi^+[K_s\to\pi^+\pi^-]] 
                               [K^{*-}\to\pi^-[K_s\to\pi^+\pi^-]]$& & --- \\
\hline
\end{tabular}
\caption{
The singly Cabibbo suppressed decays of $D$ mesons to two ground state 
are listed. Note that the notation $\pi^{(*)\pm}$ stands for $\pi^+$ or 
$\rho^+$; $\pi^{(*)0}$ stands for $\pi^0$, $\rho^0$ or $\omega^0$; 
$\phi^{(*)}$ stands for $\phi$ or $\eta^{(\prime)}$ to the extent that 
$\eta^{(\prime)}$ is an $s\overline s$ state. For each group of decays, 
we have indicated whether the tree contribution is color suppressed with 
``X'' and if it is both color and Zweig suppressed with ``XX''.
The instances which lead to an all charged final state are listed. 
The favored column are decays where the tree is colored suppressed 
and the final state has an all charged 
final state indicated by ``X''. 
Where the branching ratios are known from~\cite{pdb} we have 
included it in the last column; this is the branching ratio including the 
subsequent decays to the final all charged state indicated.
}
\label{mode_table}
\end{table}

\section{Weak Phase Determination by Initial State Mixing}
\label{phaseSection}

In this section we will consider CP violation which arises in the phase 
between $D^0$ and $\overline D^0$ decaying to a common final state $f$. 
Using the interference between the two amplitudes, the weak phase can be 
directly measured without relying on the the existence of a strong 
rescattering phase. Thus it  is useful to find new examples of CP violation 
as well as elucidating the source of CP violation in D-meson decays.

In all cases these kinds of measurements are difficult and most likely 
cannot be carried out in the near future if the weak phase is of the 
same order of magnitude of the the currently observed CP asymmetry in 
$PP$ final states.  It is possible that the small direct CP asymmetry 
seen in $D^0\to PP$ results from a large weak phase in combination with 
a small (i.e. $<O(10\%)$) strong phase. If this proves to be the case 
then a weak phase (O(10\%)) may be measured by the various methods 
considered here. A phase of this magnitude in any mode would be hard to 
explain in the SM and so would be an indication of NP.

If the CP violation in D-mesons is presumed to be from the SM source, 
the weak phase measurement can also tell us the magnitude of the penguin 
contribution. To see this, recall that the SM decay amplitude receives a 
dominant contribution from the tree which has no weak phase and a 
penguin contribution with weak phase $\gamma$. We can therefore write 
the amplitude for any SCS decay and its conjugate as:

\begin{eqnarray}
A&=&T+Pe^{+i\gamma}\nonumber\\
\overline A&=&T+Pe^{-i\gamma}\nonumber\\
\end{eqnarray}
 
\noindent
where $T$ and $P$ are generally complex numbers because they contain
strong phases. 

Thus if we know $|A|$ and $|\overline A|$ from measurement of $D^0$ and 
$\overline D^0$ decays to the final state as well as the phase 
$\theta=arg(\overline A A^*)$ then

\begin{eqnarray}
\left|\left |A\right |-\left |\overline A\right | e^{i\theta}\right|
=2|P|\sin\gamma
\end{eqnarray}

\noindent In the context of the SM, this allows us to extract the 
magnitude of the penguin, $|P|$ using the value of $\gamma$ known 
independently from the global fit to the unitarity triangle. To the 
extent that the SM contribution to the exclusive final state can be 
calculated from QCD, admittedly a very challenging task, this can be 
compared to such a measured value of $|P|$.

In any case, a precise determination of the weak phase in some modes may 
also allow us to test the SM by feeding it into the isospin analysis 
discussed in Section~\ref{isospin}. In particular, the weak phase in the 
neutral $D$ decay can allow us to determine the weak phase of the $\Delta 
I=\frac32$ channel. Such a phase would be an indication of tree-like NP.

In this section we consider three methods to accomplish a weak phase 
measurement in $D^0$ decays by producing mixed initial states of $D^0$ 
and $\overline D^0$ mesons. First we consider the ``conventional 
approach'' of taking advantage of the natural mixing of the two flavor 
eigenstates, $D^0$ versus $\overline D^0$. Since the rate of this mixing 
is small compared to the $D^0$ lifetime, we consider using the much 
larger relative mixing in the $B^0_d$ and $B^0_s$ mesons if they 
subsequently decay to a $D^0$ meson. Finally we consider methods using 
entangled $D^0\overline D^0$ states produced at tau-charm factories 
where the meson pair arises form a $\psi^{\prime\prime}$ resonance and 
also using (super) B-factories where the pair arises in a two or three 
body decay of a $B$ meson (e.g. $B\to D\overline D$ or $D\overline D K$ 
). First, however, we quickly review the oscillation formalism for 
neutral mesons:

\subsection{Oscillation Formalism}

Let us consider a generic neutral flavored meson $X$ (i.e. $X=K^0$, 
$D^0$, $B_d$ or $B_s$). Defining 
the light eigenstate ($X_L$) 
with mass $m_{XL}$ and width $\Gamma_{XL}$ and heavy eigenstate ($X_H$) 
with mass $m_{XH}$ and width $\Gamma_{XH}$, we have:

\begin{eqnarray}
\left | X_L \right \rangle
&=& 
p^X 
\left | X \right \rangle
+
q^X
\left |\overline X \right \rangle
\nonumber\\
\left | X_H \right \rangle
&=& 
p^X 
\left | X \right \rangle
-
q^X
\left |\overline X \right \rangle
\end{eqnarray}

Thus the flavor eigenstates evolve with time $t_X$ according to 

\begin{eqnarray}
\left | X(t) \right \rangle_{phys}
&=&
g^X_+ \left | X \right \rangle
-
\frac{q^X}{p^X}
g^X_- \left | \overline X \right \rangle
\nonumber\\
\left | \overline X(t) \right \rangle_{phys}
&=&
g^X_+ \left | \overline X \right \rangle
-
\frac{p^X}{q^X}
g^X_- \left | X \right \rangle
\end{eqnarray}

\noindent
where the time dependent mixing coefficients $g^X_\pm$ are given by:

\begin{eqnarray}
g^X_\pm
&=&
e^{-(im_{XH}+\frac12 \Gamma_{XH})t}
\pm 
e^{-(im_{XL}+\frac12 \Gamma_{XL})t}
\label{g:definition}
\end{eqnarray}

Let $f$ be a final state which both $X$ and $\overline X$ can decay to. 
If $A^X_f$ is the amplitude for $X\to f$ and 
$\overline A^X_f$ is the amplitude for $\overline X\to f$ then the time 
dependent rates of $X$ and $\overline X$ to $f$ are:

\begin{eqnarray}
&&\frac{d}{d\tau_X}
\Gamma(X(t)\to f)
=
\nonumber\\
&&\frac12 e^{-\tau_X}
\left [
(C^X_y+C^X_x)|A^X_f|^2
+(C^X_y-C^X_x)\left |\frac{q^X}{p^X}\overline A^X_f\right |^2
+2S^X_y Re\left ( \frac{q^{X*}}{p^{X*}}A^X_f\overline A^{X*}_f\right) 
+2S^X_x Im\left ( \frac{q^{X*}}{p^{X*}}A^X_f\overline A^{X*}_f\right )
\right ]
\nonumber\\
&&\frac{d}{d\tau_X}
\Gamma(\overline X(t)\to f)
=
\nonumber\\
&&\frac12 e^{-\tau_X}
\left [
(C^X_y+C^X_x)|\overline A^X_f|^2
+(C^X_y-C^X_x)\left |\frac{p^X}{q^X} A^X_f\right |^2
+2S^X_y Re\left ( \frac{p^{X*}}{q^{X*}}A^{X*}_f\overline A^X_f\right) 
+2S^X_x Im\left ( \frac{p^{X*}}{q^{X*}}A^{X*}_f\overline A^X_f\right )
\right ]
\label{exact_time_depend}
\end{eqnarray}

\noindent
Here 
$\Delta m_X=m_{XH}-m_{XL}$,
$\Delta \Gamma_X=\Gamma_{XH}-\Gamma_{XL}$,
$\tau_X=\Gamma_X t$, 
$x_X=\Delta m_X/\Gamma_X$, 
$y_X=\Delta \Gamma_X/(2\Gamma_X)$, 
$C^X_x=\cos(x_X\tau_X)$,
$S^X_x=\sin(x_X\tau_X)$,
$C^X_y=\cosh(y_X\tau_X)$,
$S^X_y=\sinh(y_X\tau_X)$
and $z_X=x_X-iy_X$.

In the case of $D$ mesons, both $x_D$ and $y_D\leq O(10^{-2})$,
so we will expand observables to first order in $x_D$ and 
$y_D$. 

In this limit,
the above time dependent rate 
becomes:

\begin{eqnarray}
\frac{d}{d\tau_X}
\Gamma(X(t)\to f)
=
e^{-\tau_X}
\left [
|A^X_f|^2+\tau Re\left [-iz^* A^X_f \overline 
A^{X*}_f\frac{q^*}{p^*}\right]
\right ]
+O(x^2,y^2)
\nonumber\\
\frac{d}{d\tau_X}
\Gamma(\overline X(t)\to f)
=
e^{-\tau_X}
\left [
|\overline A^X_f|^2+\tau Re\left [-iz^* A^{X*}_f 
\overline 
A^{X}_f\frac{p^*}{q^*}\right]
\right ]
+O(x^2,y^2)
\label{approx_time_depend}
\end{eqnarray}

In some of the examples below we will consider the time integrated effect 
of oscillation. To first order in $x_X$, $y_X$ this can be accomplished 
by 
replacing the decay amplitudes with ``effective'' decay amplitudes:

\begin{eqnarray}
B^X_f&=& A^X_f +\frac{i}2  \overline A^X_f \left 
(\frac{q^X}{p^X}\right)z
\nonumber\\
\overline B^X_f&=&  \overline A^X_f +\frac{i}2  A^X_f \left 
(\frac{q^X}{p^X}\right)z
\label{effective_amps}
\end{eqnarray}

\noindent
The time integrated rate for a $D^0$ meson to decay to $f$
is given, up to first order in $x_X$ and $y_X$, by using this effective 
amplitude ``without oscillation''.

Thus, for instance, 

\begin{eqnarray}
\int_0^\infty d\Gamma(X\to f)\   d\tau   &=& |B^X_f|^2 +O(x^2,y^2) 
\nonumber\\
\int_0^\infty d\Gamma(\overline X\to f)\  d\tau&=& |\overline B^X_f|^2
+O(x^2,y^2) 
\label{time_integrated}
\end{eqnarray}

\subsection{Weak Phases from $D^0$/$\overline D^0$ Oscillation}

As discussed in~\cite{Gersabeck:2011xj,Bevan:2011up}
in $D^0$ decay to a given final 
state one must consider both direct CP violation and indirect CP 
violation due to $D^0$ oscillation. Conversely, assuming that the 
oscillation parameters are known from separate studies, we can use 
oscillation extract the phase between $A^X_f$ and $\overline A^X_f$. To 
do this, it is necessary to observe the time dependence of the decays.

From eqn.~\ref{approx_time_depend} if we know $x_D$, $y_D$, $p^D$ and 
$q^D$ we see that the constant term gives the magnitudes of the 
amplitudes $|A^D_f|$ and $|\overline A^D_f|$. The slope of the decay rate 
gives the phase between these two amplitudes. If $f$ is self conjugate 
like $\pi^+\pi^-$ then such a phase difference will be CP-odd. If $f$ is 
not self conjugate, such as $\rho^+\pi^-$ then the phase will be a 
combination of CP-odd and CP-even phase differences. If both $x_D$ and 
$y_D$ are non-zero, then the phase can be determined in this way without 
ambiguity. If one of these is zero, then there is a two-fold ambiguity in 
the phase determination.

\subsection{Weak Phases from $B^0_q$/$\overline B^0_q$ Oscillation}

Another way to accomplish the measurement of the relative phase is to 
look at a two body decay of a neutral B-meson, $B_q$ for $q=d,s$, to a 
neutral D-meson where the D-meson subsequently decays to the final state 
$f$. If we observe this overall reaction $B\to M^0[D^0\to f]$ (where 
$M^0$ is a self conjugate neutral meson and $B$ is either $B_d$ or 
$B_s$) as a function of the time of the $B_q$ decay then the $D$ state 
involved in the second decay will generally be a mixture of the flavor 
eigenstates.

Of course, once the D-meson is spawned, it will oscillate as described 
above. In the following we will assume that only the B-meson decay time 
is observed and therefore the D-meson decay time is integrated over.

Let us denote by $T_{DB}$ the amplitude for $B\to M^0 D$, $T_{\overline 
D B}$ the amplitude for $B\to M^0 \overline D$, $T_{D\overline B}$ the 
amplitude for $\overline B\to M^0 D$ and $T_{\overline D\overline B}$ 
the amplitude for $\overline B\to M^0 \overline D$ and thus we can 
define the effective amplitudes for $B$ and $\overline B$ cascading down 
to the final state $f$ Using the formalism in eqn.~\ref{effective_amps} 
we can define the effective amplitudes:

\begin{eqnarray}
D^B_f&=& B^D_f T_{DB} + \overline B^D_f T_{\overline D B}
\nonumber\\
\overline D^B_f&=& B^D_f T_{D\overline B} + \overline B^D_f T_{\overline 
D \overline B}
\label{Ddef}
\end{eqnarray}

Thus the time dependent decay rate integrated over the D-meson decay time 
as a 
function of the B-meson decay time is given by 
eqn.~\ref{exact_time_depend}:

\begin{eqnarray}
&&\frac{d}{d\tau_B}
\Gamma(B(t_B)\to M [D\to f])
=
\nonumber\\
&&\frac12 e^{-\tau_B}
\left [
(C^B_y+C^B_x)|D^B_f|^2
+(C^B_y-C^B_x)\left |\frac{q^B}{p^B}\overline D^B_f\right |^2
+2S^B_y Re\left ( \frac{q^{B*}}{p^{B*}}D^B_f\overline D^{B*}_f\right) 
+2S^B_x Im\left ( \frac{q^{B*}}{p^{B*}}D^B_f\overline D^{B*}_f\right )
\right ]
\nonumber\\
&&\frac{d}{d\tau_B}
\Gamma(\overline B(t_B)\to M [D\to f])
=
\nonumber\\
&&\frac12 e^{-\tau_B}
\left [
(C^B_y+C^B_x)|\overline D^B_f|^2
+(C^B_y-C^B_x)\left |\frac{p^B}{q^B} D^B_f\right |^2
+2S^B_y Re\left ( \frac{p^{B*}}{q^{B*}}D^{B*}_f\overline D^B_f\right) 
+2S^B_x Im\left ( \frac{p^{B*}}{q^{B*}}D^{B*}_f\overline D^B_f\right )
\right ]
\label{exact_time_double_effective}
\end{eqnarray}

\noindent From the above equation, assuming that $x_B$, $y_B$ and 
$p^B/q^B$ are known then the magnitudes and relative phase of $D^B_f$ 
and $\overline D^B_f$ can be determined. Assuming that $T_{ij}$ is also 
known, then by inverting Eqn.~\ref{Ddef} we determine $B^D_f$ and 
$\overline B^D_f$. As in the last section, we can then invert the 
relations contained in Eqn.~\ref{effective_amps} to determine the 
magnitudes and relative phases of $A^D_f$ and $\overline A^D_f$. In most 
cases the amplitudes $T_{\overline DB}\approx \pm T_{D\overline B}$ will 
dominate over $T_{D B}$ and $T_{\overline D\overline B}$ and since 
$\{B^D_f,\overline B^D_f\}$, differs from $\{A^D_f,\overline A^D_f\}$, 
by $O(10^{-2})$ the phase between $D^B_f$ and $\overline D^B_f$ will, to 
a good approximation, be the negative of the phase between $A^D_f$ and 
$\overline A^D_f$.

Let us consider some particular cases of the parent $B\to MD$ decay. In 
the case of $B_d$ some candidates are $B_d\to \pi^0 \overline D^0$ 
(Br=$2.61\pm 0.24\times 10^{-4}$) and $B_d\to \rho^0 \overline D^0$ 
(Br=$3.2\pm 0.5\times 10^{-4}$). The latter is probably easier to 
observe since $\rho^0$ decays to $\pi^+\pi^-$. Indeed if the final state 
of the $D^0$ decay is either $\pi^+\pi^-$ or $\rho^0\rho^0$ then the 
entire event has an all charged final state. Other decays of this type 
are $B_d\to \eta \overline D^0$ (Br=$2.02\pm 0.35\times 10^{-4}$), 
$B_d\to \eta^\prime \overline D^0$ (Br=$1.25\pm 0.23\times 10^{-4}$) and 
$B_d\to \omega \overline D^0$ (Br=$2.59\pm 0.3\times 10^{-4}$). In 
principle the results from these modes can be combined (taking into 
account the CP of the final state). In this case we could have an 
aggregate branching ratio of $\sim 10^{-3}$. We can also consider 
$\overline D^{0*}$ instead of $\overline D^0$ and that can augment the 
effective branching ratio.

Another choice is to consider decays such as $B_d\to K_s \overline D$ 
(Br=$5.2\times 10^{-5}$) and related modes but these are an order of 
magnitude smaller in branching ratio due to Cabibbo suppression.

It is also possible to start with a $B_s$ state. The analogous decays 
are $B_s\to K_s D$ and related processes which likely have roughly the 
same branching ratios. These would include $B_s\to K_s \overline D^0$ 
and $B_s\to K^{*0} \overline D^0$ with branching ratios at the 
$10^{-4}-10^{-3}$ level (note the $K^*$ would need to decay to 
$K_s\pi^0$ to collaps the $B_s$ flavor wave function) and decays such as 
$B_s\to \phi \overline D^0$ at the $10^{-5}-10^{-4}$ level.

\subsection{Correlations at Charm and B Factories}

Let us now consider the case of a $D^0\overline D^0$ pair which is 
initially in a single, correlated, quantum state. Let us arbitrarily 
label the mesons $D_1$ and $D_2$ and consider reactions where $D_1\to 
f$; $D_2\to g$ where $f$ is the state of interest and $g$ is an 
``index'' decay (which needs to be a decay state of both $D^0$ and 
$\overline D^0$), for instance $f=\pi^+\pi^-$ and $g=K_s\pi^0$ where the 
weak phase of $\pi^+\pi^-$ is to be measured and it is assumed that the 
weak phase in $K_s\pi^0$ is small and known (i.e. just from $K_s$).

In such a scenario, the initial wave function together with the 
observation of $D_2\to g$ determines the wave function of the $D_1$ 
state as a mixture of the flavor eigenstates. In this way we are able to 
observe the interference of $D^0$ and $\overline D^0$ decay amplitudes 
to $f$~\cite{Atwood:2003mj,Atwood:2002ak}.

Starting with the wave function of the meson pair:

\begin{eqnarray}
\Psi=a 
|D_1>|\overline D_2>
+\overline a 
|\overline D_1>|D_2>
\end{eqnarray}

\noindent
where $|a|^2+|\overline a|^2=1$,
the amplitude for the combined decay $(D_1\to f)(D_1\to g)$ is therefore 

\begin{eqnarray}
A_{fg}&=&
a A_f\overline A_g
+
\overline a \overline A_f A_g
\nonumber\\
|A_{fg}|^2&=&
\frac12(|A_f|^2|\overline A_g|^2+|\overline A_f|^2|A_g|^2)
+
\frac12(|a|^2-|\overline a|^2)
(|A_f|^2|\overline A_g|^2-|\overline A_f|^2|A_g|^2)
+
2Re(a\overline a^* A_f\overline A_f^* A_g^* \overline A_g).
\label{correlated:state}
\end{eqnarray}

\noindent 
Thus if $|A_{f}|$, $|\overline A_{f}|$, $A_{g}$, $\overline A_{g}$, 
$\theta$ and $\delta$ are known, then the phase between $A_f$ and 
$\overline A_f$ can be determined.

This equation takes into account only the entanglement of the initial 
state, we can also take into account the time integrated neutral $D$ 
oscillations by integrating $|A_{fg}|^2$ to first order in $x,y$. As 
above, this is equivalent to replacing $A_{f,g}$ in 
Eqn.~\ref{correlated:state} with the effective amplitudes $B_{f,g}$ 
given by Eqn.~\ref{effective_amps}.

The conceptually simplest example 
is 
applying this at a tau-charm factory 
using the method 
of~\cite{Atwood:2002ak,Atwood:2003mj}. In this 
case, the $D^0$ pair arises from the decay of the $\psi(3770)$ and so in 
the initial state,

\begin{eqnarray}
a=+\frac{1}{\sqrt{2}}\ \ \ \ \ 
\overline a=-\frac{1}{\sqrt{2}}.
\nonumber
\end{eqnarray}

\noindent An evenly mixed $D_1$ state will thus arise when $|A_g|\approx 
|\overline A_g|$. As an example, if we take $g=K_s\pi^0$ then 
$A_g=-\overline A_g$ if, as the SM predicts, there is no CP violation in 
this pure tree decay mode except for the well understood $O(10^{-3})$ CP 
violation in the mixing of the $K_s$. (CP violation in D decay to states 
which contain $K_s$ has been observed in the related decay $D^+\to 
K_s\pi^+$ by BELLE~\cite{Ko:2012pe} and has been shown to be consistent 
with CP violation only due to the well understood mixing in the neutral 
kaon.)

This method may be generalized somewhat to the case where $g$ is a three 
body decay such as $K_s\pi^+\pi^-$. Here, the decay amplitude is a 
function of the kinematic variables. In this case we can specify the 
kinematics by the variables $E_\pm$ being the energies of the $\pi^\pm$ 
in the rest frame of the $D^0$. The amplitudes $A_g$ and $\overline A_g$ 
are functions of these variables: $A_g(E^+,E^-)$ and $\overline 
A_g(E^+,E^-)$. If the decay to $g$ is CP-invariant then the relation 
$\overline A_g(E^+,E^-)=A_g(E^-,E^+)$. If we assume that $A_g(E^+,E^-)$ 
and $\overline A_g(E^+,E^-)$ are known from other studies then 
eqn.~\ref{correlated:state} can be use to find the phase between $A_f$ 
and $\overline A_f$.

Another potential way to generate correlated neutral D-meson pairs is at 
a B factory through decays such as $B^+\to D^0 \overline D^0 K^+$ 
(Br=$2.10\pm0.26\times 10^{-3}$). More generally, it should be possible 
to adapt this analysis to the decays $B^+\to D^{*0} \overline D^0 K^+$ 
($Br=4.7\pm 1.0\times 10^{-3}$), $B^+\to D^{*0} \overline D^{*0} K^+$ 
($Br=5.3\pm 1.6\times 10^{-3}$) and $B^+\to D^0 \overline D^{*0} K^+$ 
(BR not yet known) which may increase the statistics by a factor of 
$\sim 5$.

Observing the Dalitz plot of $B^+\to D^0 \overline D^0 K^+$ decay and 
fitting it to a resonance+background will give a model for the phase of 
the decay amplitude as a function of the Dalitz plot variables.

Let us take the Dalitz plot variables to be $E_D$ and $\overline E_D$, 
the energies of the $D^0$ and $\overline D^0$ in the $B^+$ frame 
respectively so that the decay amplitude will have the dependency 
$A(E_D,\overline E_D)$. Let $E_f$ be the energy in the $B^+$ frame of 
state $f$ and $E_g$ be the energy of state $g$. The wave function of the 
D-meson pair is therefore given in terms of $E_f$ and $E_g$ by (note 
$f\leftrightarrow g$ between the two equations):

\begin{eqnarray}
a(E_f,E_g)=A(E_f,E_g)/\sqrt{|A(E_f,E_g)|^2+|A(E_g,E_f)|^2}
\nonumber\\
\overline
a(E_f,E_g)=A(E_g,E_f)/\sqrt{|A(E_f,E_g)|^2+|A(E_g,E_f)|^2}
\end{eqnarray}

\noindent This is because there is interference between the case where 
$D^0\to f$ with $\overline D^0\to g$ and $\overline D^0\to f$ with 
$D^0\to g$. Using Dalitz plot phases in this way is similar to a method 
used by BaBar to find the phase $\gamma$ in the B-meson decay to $D^0K$ 
where the D meson subsequently decays to $3\pi$~\cite{Aubert:2007ii}.

\section{Isospin Decompositions}
\label{isospin}

Since isospin is a very good symmetry of strong interactions, 
conclusions reached based on isospin alone should hold quite accurately 
in spite of some theoretical uncertainties due to hadronic interactions. 
In a recent application of isospin to D decays \cite{33b} it is argued 
that although generally isospin breaking is $O(1\%)$ the isospin 
breaking contribution to CP violation should be second order in the 
isospin breaking parameter.

The SM predicts that there is no CP violation in the $\Delta I=\frac32$ 
channel because the contribution to this channel is only through the 
tree graph $c\to d\overline d u$ while the QCD penguin which has the CP 
violating phase is pure $\Delta I=\frac12$. In principle the Electroweak 
penguin could introduce CP violation into the $\Delta I=\frac32$ channel 
but, as discussed below, this amplitude is negligibly small in $D$ 
decays.

In principle, at higher order in the SM, the electro-weak penguin (EWP) 
graphs could also contribute to CP violation in the $\Delta I=\frac32$ 
channel. We can see, however, such contributions will be very small as 
follows: First, one expects that these will be suppressed compared to 
the QCD penguin by a factor of $\alpha_W/\alpha_s\sim O(1\%)$. In the 
analogous case of $B$ physics, the EW penguins are thought to be large 
in part due to enhancement $\propto m_t$ which is not the case in charm 
decays. For example, in the decay $B\to K^+\pi^-$, $\left | A_{CP}\right
|=9\%$ so if we assume that all of this CP violation is due to EWP 
interfering with the tree graph, then we can crudely estimate the 
corresponding EWP contribution to the asymmetry in D decay as follows:

\begin{eqnarray}
A_{CP}^{EWP}
(D\to \pi^+\pi^-)
&\approx& 
\frac{\left ( EWP(D)\right )\left (Penguin(B)\right )}
{\left (Tree(D)\right ) \left ( EWP(B)\right )}
A_{CP}(B\to \pi^+ K^-)
\nonumber\\
&\approx&
\frac{Penguin(B)}{Tree(B)}
\frac{(|V_{cb}||V_{ub}|m_b)(|V_{ub}||V_{us}| )}
{(|V_{ud}||V_{cd}|)(|V_{tb}||V_{ts}|m_t)}
A_{CP}(B\to \pi^+ K^-)
\nonumber\\
&\approx&
\frac{Penguin(B)}{Tree(B)}
|V_{ub}|^2\frac{m_b}{m_t}\sim O(10^{-5})
\end{eqnarray}

\noindent which suggests even a smaller contribution. There are a number 
of channels where we can directly test the premise that isospin is a 
good symmetry in D-meson decays to two body final states. As we discuss 
below, the relative phases in decays to $\rho\rho$ provide a test of 
isospin conservation.

Turning now to the isospin decomposition of SCS D-meson decays, we 
proceed in analogy to previous work in the case of $B\to\pi\pi$ and 
related processes~\cite{Gronau:1990ka}\cite{Lipkin:1991st}. In our 
expansion we will adopt a notation similar to~\cite{Lipkin:1991st}. For 
each particular final state we will denote the isospin amplitude by 
$A^f_{\Delta I\ I}=A^f_{T:\Delta I\ I}+A^f_{P:\Delta I\ I}$ which 
indicates the amplitude for a transition through an effective 
Hamiltonian with isospin change $\Delta I$ leading to a final state of 
type $f$ with total isospin $I$. The right hand side indicates the 
further decomposition of the given amplitude into tree and penguin 
contributions respectively. Likewise the notation $A_{ij}^f =A_{T:ij}^f 
+A_{P:ij}^f$ where $i,j\in\{+-0\}$ indicates the amplitude for a decay 
with the indicated charge distribution. The corresponding amplitudes for 
$\overline D$ decay are indicated by $\overline A$.

Using this notation, we find for the $\pi\pi$ final state:

\begin{eqnarray}
A^{\pi\pi}_{+0} &=& \frac{\sqrt{3}}{2}A_{\frac32,2}^{\pi\pi}
\nonumber\\
A^{\pi\pi}_{+-} &=& 
\frac{1}{\sqrt{6}}A_{\frac32,2}^{\pi\pi}
+
\frac{1}{\sqrt{3}}A_{\frac12,0}^{\pi\pi}
\nonumber\\
A^{\pi\pi}_{00} &=& 
\frac{1}{\sqrt{3}}A_{\frac32,2}^{\pi\pi}
-
\frac{1}{\sqrt{6}}A_{\frac12,0}^{\pi\pi}
\label{iso:pipi}
\end{eqnarray}

\noindent
with the analogous relations also apply for the charge conjugate 
amplitudes.

This leads to the following ``isospin triangle'' relationships:

\begin{eqnarray}
\frac{1}{\sqrt{2}}
A^{\pi\pi}_{+-}
+
A^{\pi\pi}_{00}
-
A^{\pi\pi}_{+0}
=0
=\frac{1}{\sqrt{2}}
\overline A^{\pi\pi}_{+-}
+
\overline A^{\pi\pi}_{00}
-
\overline A^{\pi\pi}_{-0}
\label{iso:pipi:triangle}
\end{eqnarray} 

\begin{figure}
{
\includegraphics[angle=0,
width=0.6\textwidth]{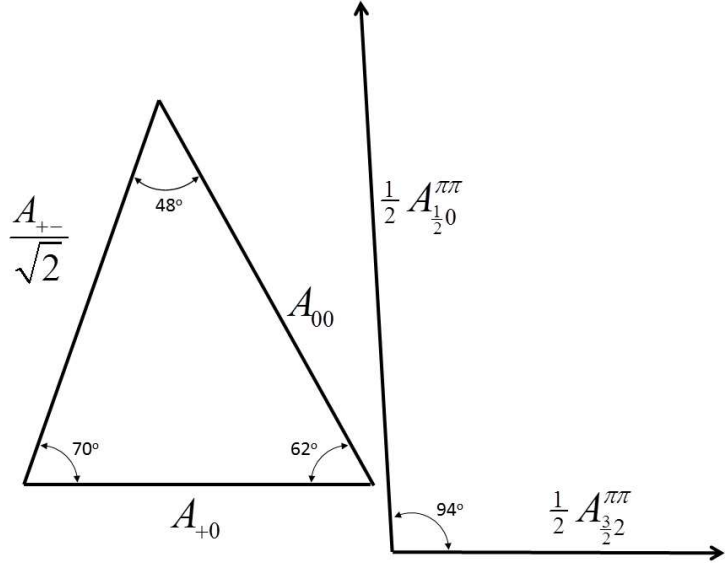}
}
\caption{
A sketch of the isospin triangle 
given in eqn.~\ref{iso:pipi:triangle}
using the central value for the branching ratios of the three modes. The 
two vectors in the sketch are proportional to the isospin 2 and isospin 0 
amplitudes as shown where the phase of the isospin 2 amplitude is 
arbitrarily taken to be real. 
}
\label{isotri}
\end{figure}

Figure~\ref{isotri} shows a sketch of such a triangle where we use 
the central values for the branching ratios involved. In the sketch we 
also show $A_{\frac12,0}^{\pi\pi}$ and $A_{\frac32,0}^{\pi\pi}$.

The case of the $KK$ final state
can also be expanded in a similar way. The resulting relations are:

\begin{eqnarray}
A^{KK}_{+0}&=&-\frac12 A^{KK}_{\frac32,1}+ A^{KK}_{\frac12,1}
\nonumber\\
A^{KK}_{+-}&=& \frac12 A^{KK}_{\frac32,1}+\frac12 A^{KK}_{\frac12,1}
+ \frac12 A^{KK}_{\frac12,0}
\nonumber\\
A^{KK}_{00}&=& 
\frac12 A^{KK}_{\frac32,1}+\frac12 A^{KK}_{\frac12,1}
-\frac12 A^{KK}_{\frac12,0}
\label{iso:KK}
\end{eqnarray}

\noindent
In this case there are three isospin amplitudes determining three decay 
amplitudes so we cannot construct a triangle relation such as 
Eqn.~\ref{iso:pipi:triangle}.

In the case of the $\pi\pi$ final state, we can see from 
Figure~\ref{isotri} that the two isospin amplitudes have a strong phase 
between them due to rescattering. In particular this illustrates the 
failure of color suppression in this system; for color suppression to be 
realized, the two amplitudes must cancel, thus be colinear in the 
complex plane and have magnitudes related by 
${\sqrt{2}}A_{\frac32,2}^{\pi\pi} = A_{\frac12,0}^{\pi\pi}$.

In contrast for the $K_sK_s$ case, evidentally there is considerable 
suppression of the branching ratio. This makes sense on the quark level 
since not only is the decay color suppressed but it is also Zweig 
suppressed. For this to happen, the isospin 1 tree amplitude and the 
isospin 0 tree amplitude must cancel fairly well: $ \frac12 
A^{KK}_{T:\frac32,1}+\frac12 A^{KK}_{T:\frac12,1} \approx \frac12 
A^{KK}_{T:\frac12,0}$. Since the dominant tree amplitude is suppressed, 
this suggests the possibility that $A_{CP}$ in $D^0\to K_sK_s$ could be 
enhanced compared to the $K^+K^-$ case at the expense of the total rate.

The $\rho\rho$ final states have the same form as $\pi\pi$ except that 
there are three polarization states which can arise from the decay of a 
scalar, $A_\parallel$, $A_\perp$ and $A_\ell$. There will thus be a 
separate set of isospin amplitudes so the analog to the above 
decomposition is:

\begin{eqnarray}
A^{\rho\rho(i)}_{+0} &=& \frac{\sqrt{3}}{2}A_{\frac32,2}^{\rho\rho(i)}
\nonumber\\
A^{\rho\rho(i)}_{+-} &=& 
\frac{1}{\sqrt{6}}A_{\frac32,2}^{\rho\rho(i)}
+
\frac{1}{\sqrt{3}}A_{\frac12,0}^{\rho\rho(i)}
\nonumber\\
A^{\rho\rho(i)}_{00} &=& 
\frac{1}{\sqrt{3}}A_{\frac32,2}^{\rho\rho(i)}
-
\frac{1}{\sqrt{6}}A_{\frac12,0}^{\rho\rho(i)}
\label{iso:rhorho}
\end{eqnarray}

\noindent
where $i\in\{\perp,\parallel,\ell\}$ indexes the polarization state. 

For each polarization state we therefore also have an isospin triangle 
relation:

\begin{eqnarray}
\frac{1}{\sqrt{2}}
A^{\rho\rho(i)}_{+-}
+
A^{\rho\rho(i)}_{00}
-
A^{\rho\rho(i)}_{+0}
=0
=\frac{1}{\sqrt{2}}
\overline A^{\rho\rho(i)}_{+-}
+
\overline A^{\rho\rho(i)}_{00}
-
\overline A^{\rho\rho(i)}_{-0}
\label{iso:rhorho:triangle}
\end{eqnarray} 

\noindent For each $\rho\rho$ final state, an angular 
analysis~\cite{Chiang:1999qn} provides the magnitude of the three 
polarization amplitudes and the cosine of the phase angle between them. 
Thus (up to a 2 fold ambiguity) the relative phases between these three 
amplitudes can be determined. The relation 
(Eqn.~\ref{iso:rhorho:triangle}) gives the phase between the three 
amplitudes with the same charge distribution in terms of their 
magnitudes (again up to a 2 fold ambiguity). Combining the two kinds of 
information, we have 18 phase differences for 9 amplitudes in D-decays 
(Taking into account an overall phase, the system is overdetermined by 8 
degrees of freedom) which checks the validity of 
Eqn.~\ref{iso:rhorho:triangle}. Since this relation was derived using 
isospin conservation, the validity of this symmetry is thus quantified.

For the $\rho\pi$ final state the decomposition is:

\begin{eqnarray}
A^{\rho\pi}_{+0} &=& 
\frac{\sqrt{3}}{\sqrt{8}}A_{\frac32,2}^{\rho\pi}
-\frac{1}{\sqrt{8}}A_{\frac32,1}^{\rho\pi}
+\frac{1}{\sqrt{2}}A_{\frac12,1}^{\rho\pi}
\nonumber\\
A^{\rho\pi}_{0+} &=& 
\frac{\sqrt{3}}{\sqrt{8}}A_{\frac32,2}^{\rho\pi}
+\frac{1}{\sqrt{8}}A_{\frac32,1}^{\rho\pi}
-\frac{1}{\sqrt{2}}A_{\frac12,1}^{\rho\pi}
\nonumber\\
A^{\rho\pi}_{+-} &=& 
\frac{1}{\sqrt{12}}A_{\frac32,2}^{\rho\pi}
+\frac{1}{2}A_{\frac32,1}^{\rho\pi}
+\frac{1}{2}A_{\frac12,1}^{\rho\pi}
+\frac{1}{\sqrt{6}}A_{\frac12,0}^{\rho\pi}
\nonumber\\
A^{\rho\pi}_{-+} &=& 
\frac{1}{\sqrt{12}}A_{\frac32,2}^{\rho\pi}
-\frac{1}{2}A_{\frac32,1}^{\rho\pi}
-\frac{1}{2}A_{\frac12,1}^{\rho\pi}
+\frac{1}{\sqrt{6}}A_{\frac12,0}^{\rho\pi}
\nonumber\\
A^{\rho\pi}_{00} &=& 
\frac{1}{\sqrt{3}}A_{\frac32,2}^{\rho\pi}
-\frac{1}{\sqrt{6}}A_{\frac12,0}^{\rho\pi}
\end{eqnarray}

\noindent
which in turn leads to the following pentagonal isospin relationships:

\begin{eqnarray}
\sqrt{3}\  A^{\rho\pi}_{\frac32,2}
&=&
\sqrt{2}(A^{\rho\pi}_{+0}+A^{\rho\pi}_{0+})
=A^{\rho\pi}_{+-}+A^{\rho\pi}_{-+}+2A^{\rho\pi}_{00}
\nonumber\\
\sqrt{3}\  \overline A^{\rho\pi}_{\frac32,2}
&=&
\sqrt{2}(\overline A^{\rho\pi}_{-0}+\overline A^{\rho\pi}_{0-})
=\overline A^{\rho\pi}_{-+}+\overline A^{\rho\pi}_{+-}
+2\overline A^{\rho\pi}_{00}
\label{iso:rhopi:pentagon}
\end{eqnarray}

\noindent
also the $\Delta I=3/2$ contribution to the $I=1$ final state 
follows from the relation:

\begin{eqnarray}
3\  A^{\rho\pi}_{\frac32,1}
&=& \sqrt{2}(A^{\rho\pi}_{0+}-A^{\rho\pi}_{+0})
+2(A^{\rho\pi}_{+-}-A^{\rho\pi}_{-+})
\nonumber\\
3\  \overline A^{\rho\pi}_{\frac32,1}
&=& \sqrt{2}(\overline A^{\rho\pi}_{0+}-\overline A^{\rho\pi}_{+0})
+2(\overline A^{\rho\pi}_{+-}-\overline A^{\rho\pi}_{-+})
\label{iso:rhopi:pentagon2}
\end{eqnarray}

In the case of the decay $D_s\to\pi K^*$ the isospin decomposition of the 
amplitudes is:

\begin{eqnarray}
A^{\pi K^*}_{+0}
&=&
\frac{1}{\sqrt{3}}A^{\pi K^*}_{\frac32}+
\frac{\sqrt{2}}{\sqrt{3}}A^{\pi K^*}_{\frac12}
\nonumber\\
A^{\pi K^*}_{0+}
&=&
\frac{\sqrt{2}}{\sqrt{3}}
A^{\pi K^*}_{\frac32}-
\frac{1}{\sqrt{3}}A^{\pi K^*}_{\frac12}
\nonumber\\
{\rm Thus, \ \ \ \ }
\sqrt{3}A^{\pi K^*}_{\frac32}
&=&
A^{\pi K^*}_{+0}
+\sqrt{2}
A^{\pi K^*}_{0+}
\label{Ds:relations}
\end{eqnarray}

\noindent
In this case, the two decay amplitudes depend on two isospin amplitudes
so there is no isospin triangle relation as in the case of $\pi\pi$ and 
$\rho\pi$.

\subsection{Phases in Dalitz Plots}

In the two body decays $D\to\rho\pi$ and $D_s\to K^*\pi$ the vectors 
decay in turn to two pseudoscalars, $\rho\to\pi\pi$ and $K^*\to K\pi$. 
The final states are therefore three body Dalitz decays\cite{ads,ggsz}. 
The same three scalar final state will, in general, receive 
contributions from a number of different pseudo two body channels. For 
example in the case of $D_s\to K^*\pi$, the two charge distributions 
will contribute to the same three body final state, in particular 
$D_s\to K^{*+}\pi^0\to K^0\pi^+\pi^0$ and $D_s\to K^{*0}\pi^+\to 
K^0\pi^0\pi^+$. Thus the $K^0\pi^0\pi^+$ final state receives 
contributions from both the $K^{*0}\pi^+$ and $K^{*+}\pi^0$ channels. A 
fit to the the distribution in the Dalitz plot variables will therefore 
determine both the magnitudes of the two body amplitudes and also the 
relative phase between them as well as other channels which contribute 
to this final state such as $K^0\rho^+$. Note that the other decay of 
the $K^*$ in the above will not involve interference between these two 
channels, in particular $D_s\to K^{*+}\pi^0\to K^+\pi^0\pi^0$ and 
$D_s\to K^{*0}\pi^+\to K^+\pi^-\pi^+$.
 
The same situation also applies to $D^0\to\rho\pi$ which leads to the 
final state $\pi^+\pi^-\pi^0$. In this case the pseudo two body channels 
$\rho^0\pi^0$, $\rho^+\pi^-$ and $\rho^-\pi^+$ all contribute so in 
fitting the Dalitz plot one obtains the magnitude and relative phases of 
each of these channels.

\subsection{Standard Model Tests using Isospin}

The main test of the SM origin for CP violation in hadronic $D$ decays 
which can be accomplished using isospin analysis is to test the SM 
prediction that the tree graph which is the only contribution to the 
$\Delta I=3/2$ Hamiltonian, has no phase in the Wolfenstein phase 
convention. Thus, assuming EWP are negligible, any CP violation in phase 
or magnitude is contained in the $\Delta I=1/2$ component which receives 
contributions both from the tree and the penguin.

In each system of decays there are therefore two kinds of tests which,
 in 
principle, can be performed. 
\begin{enumerate}
\item
The magnitude of the 
$\Delta I=3/2$ transition amplitude is the same for the decay and its 
charge conjugate. 
\item
The phase of the 
$\Delta I=3/2$ transition amplitude is the same for the decay and its 
charge conjugate. 
\end{enumerate}

In the $\pi\pi$ final state, the system is sufficiently simple to allow 
us to cleanly extract three isospin related CP asymmetries. To fully 
characterize CP violation in this system, a fourth quantity must be 
determined by a phase measurement of the type described in 
Section~\ref{phaseSection}.

Let us denote by $\delta^{f}_{ij}$ the partial rate difference for final 
state $f$ with charge distribution $ij$, i.e. 
$\delta^{f}_{ij}=|A^{f}_{ij}|^2-|\overline A^{f}_{ij}|^2$. 
Likewise for 
the isospin amplitudes denote:

\begin{eqnarray}
\delta^{f}_{\Delta I \ I}
&=&|A^{f}_{\Delta I \ I}|^2- |\overline A^{f}_{\Delta I \ I}|^2
\nonumber\\
\delta^{f}_{[\Delta I \ I;\Delta J \ J]}
&=& Re\left (
A^{f}_{\Delta I \ I}A^{f*}_{\Delta J \ J}
-
\overline A^{f}_{\Delta I \ I}\overline A^{f*}_{\Delta J \ J}
\right )
\end{eqnarray}

\noindent Using this notation, Eqn.~\ref{iso:pipi} implies that the CP 
violation in the $D\to\pi\pi$ decays can be rewritten as:

\begin{eqnarray}
\delta^{\pi\pi}_{+0}
&=&\frac34 
\delta^{\pi\pi}_{\frac32 2}
\nonumber\\
\delta^{\pi\pi}_{+-}
&=&
\frac16 \delta^{\pi\pi}_{\frac32 2}
+
\frac13 
\delta^{\pi\pi}_{\frac12 0}
+
\frac13\sqrt{2}
\delta^{\pi\pi}_{\frac32 2;\frac12 0}
\nonumber\\
\delta^{\pi\pi}_{00}
&=&
\frac13 \delta^{\pi\pi}_{\frac32 2}
+
\frac16 
\delta^{\pi\pi}_{\frac12 0}
-
\frac13\sqrt{2}
\delta^{\pi\pi}_{\frac32 2;\frac12 0}
\end{eqnarray}

\noindent
Since this gives each of the observed partial rate differences in terms 
of three different CP violating underlying isospin quantities, we can 
invert these and obtain:

\begin{eqnarray}
\delta^{\pi\pi}_{\frac32 2}
&=&\frac43 
\delta^{\pi\pi}_{+0}
\nonumber\\
\delta^{\pi\pi}_{\frac12 0}
&=&
2
\delta^{\pi\pi}_{+-}
+
2\delta^{\pi\pi}_{00}
-
\frac43
\delta^{\pi\pi}_{+0}
\nonumber\\
\frac{1}{\sqrt{2}}
\delta^{\pi\pi}_{\frac32 2;\frac12 0}
&=&
\frac13\delta^{\pi\pi}_{+0}
+
\frac12\delta^{\pi\pi}_{+-}
-
\frac12\delta^{\pi\pi}_{00}
\end{eqnarray}

\noindent 
As discussed in ~\cite{33b}
the first expression for $\delta^{\pi\pi}_{\frac32 2}$ 
implies, $\delta^{\pi\pi}_{\frac32 2}=0$ is a test of type (1) due to the 
evident fact that the decay to $\pi^+\pi^0$ is governed only by the 
$\Delta I=\frac32$ Hamiltonian.

The other two combinations indicate different features of CP violation 
in the $\Delta I=\frac12$ channel which could be entirely due to SM 
physics. As discussed in~\cite{33b}, for $\delta^{\pi\pi}_{\frac12 0}$ 
to be non-zero requires that there are two contributions to this isospin 
channel which have different strong phases and also different weak 
phases. This would generally be expected to be the case in the SM since 
both tree and penguin contribute to $\Delta I=\frac12$. It could, 
however, happen that $\delta^{\pi\pi}_{\frac12 0}$ is small due to the 
strong phase difference between the two contributions to $\Delta 
I=\frac12$ being small but in this case the quantity 
$\delta^{\pi\pi}_{\frac32 2;\frac12 0}$ could be non-zero due to strong 
and weak phase difference between the two different isospin channels.

To make this clear, consider, for example, what happens if 
$\delta^{\pi\pi}_{\frac32 2}=\delta^{\pi\pi}_{\frac12 0}=0$ but 
$\delta^{\pi\pi}_{\frac32 2;\frac12 0}\neq 0$. This would imply first 
that $|A^{\pi\pi}_{\frac32 2}|=
|\overline A^{\pi\pi}_{\frac32 2}|$ 
and
$|A^{\pi\pi}_{\frac12 2}|= 
|\overline A^{\pi\pi}_{\frac12 2}|$ 
but that the phase between 
$A^{\pi\pi}_{\frac32 2}$
and
$A^{\pi\pi}_{\frac12 0}$
is different than the phase between 
$\overline A^{\pi\pi}_{\frac32 2}$
and
$\overline A^{\pi\pi}_{\frac12 0}$
resulting from a different weak phase between the two isospin 
channels.

The measurement of the phase difference between either of the neutral 
amplitudes and their charge conjugates, i.e. $A^{\pi\pi}_{+-}$ versus 
$\overline A^{\pi\pi}_{+-}$ or $A^{\pi\pi}_{00}$ versus $\overline 
A^{\pi\pi}_{00}$ using the methods in Section~\ref{phaseSection} allows 
the complete determination of all the amplitudes and therefore all the 
CP violation in this system. This follows from relation 
eqn.~\ref{iso:pipi:triangle} which implies that the three amplitudes 
form a triangle in the complex plane. The phase between 
$A^{\pi\pi}_{+0}$ and either of the neutral modes is therefore 
determined (up to a 2 fold ambiguity) and so the phase of 
$A^{\pi\pi}_{+0}$ is known. The same is also true for the charge 
conjugate amplitudes so ultimately the weak phase between 
$A^{\pi\pi}_{+0}$ and $\overline A^{\pi\pi}_{-0}$ is determined (up to a 
four fold ambiguity). This then is a test of the SM of type (2).

Note that if the phase difference with the conjugates is measured for 
both $\pi^+\pi^-$ and $\pi^0\pi^0$ final states, then there is a 
consistency check for the isospin relations because the isospin triangle 
for $D^0$ decay fixes the phase between $A^{\pi\pi}_{+-}$ and 
$A^{\pi\pi}_{00}$ while the the isospin triangle for $\bar D^0$ decay 
fixes the phase between $\overline A^{\pi\pi}_{+-}$ and $\overline 
A^{\pi\pi}_{00}$. In addition, having both phase measurements will 
resolve the four fold ambiguity with respect to the orientation of the 
isospin triangles. Of course measuring the weak phase directly with the 
$\pi^0\pi^0$ final state using the methods described above will likely be 
experimentally difficult.

The same discussion also applies to each polarization of the final state. 
the $\rho\rho$ final state. Because all of the relative phases of the 9 
$D\to\rho\rho$ amplitudes can be measured as discussed above (and 
likewise for the 9 $\overline D\to\rho\rho$ amplitudes), if one weak 
phase measurement is made then all of the weak phase differences are 
known. In principle, there are six possible weak phase differences (2 
modes $\times$ 3 polarizations) which can be measured in $D^0$ decay to 
$\rho^0\rho^0$ or $\rho^+\rho^-$ so there are multiple checks on this 
kind of measurement.

For the $\rho\rho$ final state then we have 3 type (1) tests of the SM by 
comparing the magnitude of  each of the $\rho^+\rho^0$ amplitudes with 
their conjugates. There are two other independent tests which can be made 
by comparing the phase differences between the amplitudes with the phase 
differences of their conjugates. Finally, with an absolute weak phase 
determination and the isospin relationships Eqn.~\ref{iso:rhorho} we can 
have 3 type (2) tests for the weak phase difference between each of the 
$\rho^+\rho^0$ polarizations and their conjugates.

In the case of $D\to\rho\pi$ the phase between the three $D^0\to \rho 
\pi$ amplitudes can be determined from analysis for the Dalitz plot 
distributions of $D^0\to\pi^+\pi^-\pi^0$. These amplitudes are therefore 
a part of a more general isospin analysis of $D\to 3\pi$ as considered 
in~\cite{Gaspero:2008rs}. Using the relationship 
eqn.~\ref{iso:rhopi:pentagon} we can see that $A^{\rho\pi}_{\frac32,2}$ 
is determined as a linear combination of these three amplitudes and so is 
determined up to an overall weak phase. Likewise we can extract the 
charge conjugate so the SM can be tested by comparison of 
$|A^{\rho\pi}_{\frac32,2}|$ with $|\overline A^{\rho\pi}_{\frac32,2}|$.

Furthermore, the relation eqn.~\ref{iso:rhopi:pentagon} gives 
$A^{\rho\pi}_{\frac32,2}$ as a linear combination of the related charged 
D-meson decays $A^{\rho\pi}_{+0}$ and $A^{\rho\pi}_{0+}$ so that the 
phase of these two decays relative to the neutral decays can be 
determined up to a two fold ambiguity. We can thus use 
eqn.~\ref{iso:rhopi:pentagon2} to find the magnitude of 
$|A^{\rho\pi}_{\frac32,1}|$. Likewise we can extract the charge conjugate 
of the same amplitude and so SM can be tested by comparison of 
$|A^{\rho\pi}_{\frac32,1}|$ with $|\overline A^{\rho\pi}_{\frac32,1}|$.
Thus we have two tests of type (1) in this system.

We can generate the corresponding type (2) tests for both of the $\Delta 
I=3/2$ amplitudes in $D\to\rho\pi$, if we know the weak phase difference 
between at least one of the neutral modes and its conjugate
using the methods of Section~\ref{phaseSection}. Since the relative 
phases between all the D decays are determined by the construction above, 
the weak phase difference will then be determined. The weak phases of the 
other two neutral cases (all of which are found in the same Dalitz plot) 
would then provide consistency checks.

In the case of $D_s\to K^*\pi$ there is just a SM check of type (1). In 
this case, the phase between the two amplitudes $A^{\pi K^*}_{+0}$ and 
$A^{\pi K^*}_{0+}$ can be determined from the $K^+\pi^-\pi^0$ Dalitz 
plot. Thus using eqn.~\ref{Ds:relations} we obtain the magnitude of 
$|A^{\pi K^*}_{\frac32}|$. Again we can test the SM through verifying 
$|A^{\pi K^*}_{\frac32}|=|\overline A^{\pi K^*}_{\frac32}|$. Unlike the 
above cases, there is no way to determine the weak phase of this 
amplitude because there is no neutral decay related by isospin.

\section{General Requirements for Testing CP violation in SCS decays of 
D-mesons}
\label{numerics}

In order to form a rough estimate of the requirements to find CP 
violation and test the SM through the modes above let us assume that the 
CP violation in these SCS modes is generally at the same level as seen in 
the SCS modes (e.g. $\pi\pi$ and $KK$), 
on the order of $0.1-1\%$. In terms of raw 
statistics, a sample of $10^5-10^7$ would be required. Since the 
branching ratios of these modes is typically $10^{-3}$ this would mean 
that $10^8-10^{10}$ D-mesons would be required; and probably an order of 
magnitude more depending on the acceptance for various decay modes. 
Indeed this is roughly true in the LHCb results~\cite{Aaij:2011in} based 
on an integrated luminosity of $0.62fb^{-1}$ the yield of $K^+K^-$ was 
$1.44\times 10^6$ and the yield of $\pi^+\pi^-$ was $0.38\times 10^6$. 
These results point out important challenges which must be overcome to 
carry out such studies at LHCb and more generally at other facillities.

At the LHCb it is, of course crucial to overcome the fact that the 
initial state is not charge conjugate. This, of course, is less of a 
problem at $e^+e^-$ colliders such as B-factories or tau-charm factories. 
In any case, aside from the requirement of raw statistics, it is 
necessary to identify and tag the initial D-meson and find the 
various final states. To this end, as discussed in 
Section~\ref{candidates}, final states with all charged final state 
particles (i.e. $\pi^\pm$ and $K^\pm$) will be easier to detect.

Determining the phase through any of the methods discussed in 
section~\ref{phaseSection} may require statistics somewhat beyond 
currently planned facilities. First consider using straight D 
oscillation with Eqn.~\ref{approx_time_depend}. Obviously the first 
requirement is the ability to track the time dependence of the decay to a 
precision $<< 1/\Gamma_{D^0}$. The relevant terms in the time dependence 
which we need to extract is the term $\propto\tau$. This term, of course, 
is multiplied by the relative rate of mixing $|z|\sim 10^{-2}$. 
Furthermore, if the weak phase is similar to the observed level of CP 
violation in magnitude for $\pi^+\pi^-/K^+K^-$ then we would expect 
$arg(A\overline A^*)\sim 1\%$. If this is indeed the case you would need 
$10^9$ final states in order to see the decay and, since the branching 
ratio is $10^{-3}$ you would therefore need $\sim 10^{12}$ D-mesons to 
start with.

Using the double oscillation method, for example 
eqn.~\ref{exact_time_double_effective} in the $B_d$ case where $y$ is 
small and $p/q=e^{2i\beta}$, the relevant term would be the one 
proportional to $S^B_x$. If there was no weak phase then this would be 
proportional to the same $\sin 2\beta$ as $B\to\psi K_s$. In effect then, 
we would be looking for a deviation from the SM value of this coefficient 
by $O(1\%)$ so we would expect to need $\sim 10^5$ decays to preform the 
measurement. In the case of the $\pi\pi$ final state, the combined 
branching ratio would be $4.2\times 10^{-7}$ for the channel through 
either $D^0\pi^0$ or $D^0\rho^0$. This gives an initial requirement for 
the 
number of B-mesons to be $\sim 2\times 10^{11}$. By combining a number of 
modes (e.g. $B^0\to \overline D^0 \pi$, $B^0\to \overline D^0 \rho$, 
$B^0\to \overline D^{0*} \pi$ etc.) it may be possible to reduce this to 
the $10^{10}-10^{11}$ range.

Using the correlation method, if we take the decay $B^+\to K^+ 
D^0\overline D^0$ and use the index decay $g=K_s\pi^0$ with the final 
state $f=\pi^+\pi^-$ and assuming we need to observe $10^5$ events, then, 
not including acceptance, the numbers of $B$ mesons needed is $3\times 
10^{12}$. If we broaden the method to include $B^+\to K^+ D^{*0}\overline 
D^{*0}$ (assuming a total Br=1\%) and use as an index state 
$g=K_s\pi^+\pi^-$ and a target state $f=\rho^0\rho^0$, the number is 
reduced to $1.5\times 10^{11}$.

Using correlations at a $\psi^{\prime\prime}$ factor, the number of $DD$ 
pairs required using the above assumptions with index state 
$g=K_s\pi^+\pi^-$ and a target state $f=\rho^0\rho^0$ is $5\times 10^9$.

It seems then that each of these methods requires an input of $\sim 
10^{11}$ mesons if the phase is of the same order of magnitude as 
$A_{CP}$ for $\pi\pi$ and $KK$. 
This is probably beyond the capability of machines in the 
foreseeable future. If, however, the CP violating phase is an order of 
magnitude larger than $A_{CP}$ (i.e. because the strong phase was 
O(10\%)) then these requirements would be reduced by 2 orders of 
magnitude and perhaps such experiments could become possible at super B 
factories or the LHCb upgrades. 

Perhaps the cleanest environment to measure such 
phases would be at high luminosity charm factories where $\sim 10^{10}$ 
meson pairs would be needed if the phase is O(1\%). Again if the phase 
were 10\% this would be reduced by two orders of magnitude.

\section{Summary and Conclusion}
\label{conclusion}

$D^0$ mixing is unique as its the only charge 2/3 bound system providing 
us with a great opportunity to search for new physics. In many 
interesting BSM scenarios enhanced mixing and also enhanced CP 
asymmetries are expected; warped extra dimension models are a well known 
example. The recent discovery of direct CP violation in $D^0$-decays by 
the LHCb collaboration gives a huge impetus to these searches.  The 
observed CP asymmetry of O($0.5\%$) is somewhat bigger than some 
estimates though it seems SM explanation is quite plausible. Hadronic 
uncertainties make precise predictions exceedingly difficult, therefore, 
for now, possible role of new physics cannot be ruled out. More 
experimental information may well be pivotal in this instance This is 
the basic rationale behind this work leading us to make several 
suggestions.

We suggest that the observed enhanced effects due to non-perturbative 
physics may be most pronounced for the exclusive two pseudoscalar modes 
only, {\it e.g.} $\pi \pi$ and $K \overline K$. For multipaticle 
(inclusive) final states, the quark level CP asymmetry of about 6 
$\times 10^{-4}$ may be releava nt. A simple way to implement this 
experimentally may be to look for (say), decays of $D$ to final states 
with a $K$ and a $\overline K$ where the sum of their energies is less 
than the energy of the parent D. If these inclusive final states also 
show enhanced CP asymmetries (say at the level seen in exclusive $K^+ 
K^-$, $\pi^+ \bar \pi^-$), then that would mean that it has a new 
physics origin otherwise it will give support to a SM explanation.

Since the tree contribution is likely suppressed in color-suppressed 
final states, it is likely that CP asymmetries will be enhanced therein. 
To fecilitate experimental detection final states leading to charged 
$\pi$s may be best to focus on. These twin considerations lead us to 
suggest $D^0 \to \rho^0 \rho^0$, $D_s \to \rho K^+$ and $D_s \to \rho^0 
K^{*+}$ especially interesting. The vector vector final states have the 
additional bonus that angular correlations can also be used for 
additional CP-violating observables.

The importance of CPT constraints on CP violating observables are 
emphasized and illustrated with regard to exlusive and inclusive and 
radiative modes.

While SU(3) and Uspin symmetries seem quite badly broken in D decays, 
isospin likely holds quite well motivating us to to investigate its use 
especially in decays such as $D \to \pi \pi$, $\rho \pi$, $\rho \rho$ as 
well as for $D_s \to K^* \pi$.

We also studied how such analysis may be augmented by information about 
the weak phase in $D^0$ decays. To do this it is necessary to study a 
sample of D-mesons which are in a mixed state of $D^0$ and $\overline 
D^0$. Such a state may result from $D^0\overline D^0$ oscillation or 
from the decay of a $B$ or $B_s$ meson which itself is in a mixed state 
due to its oscillation. Alternatively, if a $D^0\overline D^0$ pair is 
in an entangled state, the observation of the decay of one neutral 
D-meson implies the other D-meson is in a mixed state. Such entangled 
pairs may be produced in charm factories through the decay of 
$\psi^{''}$ or as the result of B-meson decays.

\section*{Note Added I}
Since the role of hadronic matrix elements of penguin operators in charm CP has been of considerable discussions and speculations,
we take this opportunity to briefly draw attention to a recent work on the lattice (Ref~\cite{RBC-UKQCD12})  of the RBC and UKQCD collaborations,
on the origin of the large enhancement in $K \to \pi \pi$ decays in the $I=0$ final state relative to $I=2$
that often goes under the name of the ``$\Delta I=1/2$ puzzle". That work finds that, at a renormalization
scale of around 1.7 GeV or more,  the entire enhancement originates from non-perturbative matrix elements
of simple tree operators ({\it i.e.} $Q_1$ and $Q_2$) and the contribution of the penguin operators is quite negligible. 

\section*{Note Added II}
We want to take the opportunity to briefly mention two new experimental results from LHCb.
In  LHCb-CONF-2013-003  they give  a new preliminary result using soft pion technique as in ~\cite{Aaij:2011in}
but now with 1.0 $fb^{-1}$ of data; they find  $\Delta A_{CP} =( -0.34 \pm 0.15 \pm 0.10) \%$.
On the ther hand, using B-semileptonic tags on the same  amount of data, they report ~\cite{Bsemi},
 $\Delta A_{CP} =( +0.49 \pm 0.30 \pm 0.14) \%$, yielding a new world average,  $\Delta A_{CP} =( -0.33 \pm 0.12) \%$~\cite{TGBF13}.

\section*{Acknowledgements}

The work of AS was supported in part by the U.S. DOE contract 
\#DE-AC02-98CH10886(BNL). The work of DA was supported in part by the 
U.S. DOE contract \#DE-FG02-94ER40817 (ISU).

\end{document}